\documentclass[twocolumn, 
reprint,
superscriptaddress,
preprintnumbers,
showpacs,
amsmath,amssymb,
prb]{revtex4-2}

\usepackage{graphicx}
\usepackage[font={small,it},justification=justified,format=plain]{caption}
\usepackage{dcolumn}
\usepackage{bm}
\usepackage{siunitx}
\usepackage{subfig}
\usepackage{listings}
\usepackage[dvipsnames, table]{xcolor}
\usepackage{float}
\usepackage{wrapfig}
\usepackage{dsfont}
\setcitestyle{super}
\usepackage{hyperref}
\hypersetup{colorlinks=true,allcolors=MidnightBlue}
\usepackage{xcolor}
\usepackage{graphicx, nicefrac}
\usepackage{multirow}
\usepackage{float}
\usepackage{mathtools}
\usepackage{colortbl} 
\usepackage{booktabs}
\usepackage{boldline}
\usepackage{hhline}
\usepackage{makecell}
\usepackage{array}

\definecolor{cloud}{HTML}{A3AFBB}
\definecolor{slate}{HTML}{7493A3}
\definecolor{navy}{HTML}{395464}
\definecolor{sea}{HTML}{668A82}
\definecolor{tan}{HTML}{D8CAAC}
\definecolor{copper}{HTML}{B68044}
\definecolor{lightgrey}{HTML}{EDEDED}

\newcommand{\rslate}{\rowcolor[HTML]{6E8098}}
\newcommand{\tcw}{\color{white}}
\newcommand{\tcb}{\color{black}}
\newcommand{\seacl}{\cellcolor{sea}}
\newcommand{\spc}{\phantom{~}}
\newcommand{\sss}{\scriptscriptstyle}
\newcommand{\overhang}{\dimexpr\tabcolsep+0.1pt\relax}
\newcommand{\tbl}{\cellcolor[HTML]{EEF2F4}}
\newcommand{\tbd}{\cellcolor[HTML]{DBE3E8}}
\newcommand\edit[1]{\textcolor{black}{#1}}
\newcommand\editt[1]{\textcolor{black}{#1}}

\newcommand*{\citen}[1]{
  \begingroup
    \romannumeral-`\x
    \setcitestyle{numbers}
    \cite{#1}
  \endgroup   
}

\begin{document}

\title{Optimization strategies developed on NiO for Heisenberg exchange coupling calculations using projector augmented wave based first-principles DFT+U+J}

\author{L\'{o}rien MacEnulty}
 \email{lmacenul@tcd.ie}
\author{David D. O'Regan}
\affiliation{
 School of Physics, Trinity College Dublin, The University of Dublin, Dublin 2, Ireland\\
}

\date{\today}

\begin{abstract}
High-performance batteries, heterogeneous catalysts and next-generation photovoltaics often centrally involve transition metal oxides (TMOs) that undergo charge or spin-state changes. Demand for accurate DFT modeling of TMOs has increased in recent years, driving improved quantification and correction schemes for approximate DFT’s characteristic errors, notably those pertaining to self-interaction and static correlation. Of considerable interest, meanwhile, is the use of DFT-accessible quantities to compute parameters of coarse-grained models such as for magnetism. To understand the interference of error corrections and model mappings, we probe the prototypical Mott-Hubbard insulator NiO, calculating its electronic structure in its antiferromagnetic I/II and ferromagnetic states. We examine the pronounced sensitivity of the first principles calculated Hubbard \editt{parameters U and J}  to choices concerning Projector Augmented Wave (PAW) based population analysis, we reevaluate spin quantification conventions for the Heisenberg model, and we seek to develop best practices for calculating Hubbard parameters specific to energetically meta-stable magnetic orderings of TMOs. Within this framework, we assess several corrective functionals using in situ calculated U and J parameters, e.g., DFT+U and DFT+U+J. We find that while using a straightforward workflow with minimal empiricism, the NiO Heisenberg parameter RMS error with respect to experiment was reduced to 13\%, an advance upon the state-of-the-art. Methodologically, we used a linear-response implementation for calculating the Hubbard U available in the open-source plane-wave DFT code \textsc{Abinit}. We have extended its utility to calculate the \editt{intra-atomic} exchange coupling J, however our findings are anticipated to be applicable to any DFT+U implementation.

\begin{description}
\item[Keywords]
\it{DFT+U, DFT+U+J, PAW, nickel oxide, Heisenberg exchange}
\end{description}
\end{abstract}

\maketitle

\section{\label{Introduction}Introduction}

Antiferromagnetic transition metal oxides (TMOs) are increasingly vital constituents of many technologies, from high-performance battery cathodes, \cite{chakraborty_accurate_2018,saritas_charge_2020,shu_comparison_2010,meng_first_2009,shishkin_redox_2021} to water-splitting catalysts,\cite{hegner_database_2016,azizthesis} to efficient photovoltaic devices for sustainable energy storage.\cite{Siebentritt2012,Polman2016,Yan2017} Understanding the macroscopic magnetic properties of this class of material requires a sophisticated description of the competing spin-related energy scales relevant at  microscopic scales.

\edit{A variety of simulation techniques exist to calculate the material properties of TMOs, such as DMFT,\cite{Zhang2019,kvashnin2015,mandal2019,kunes2007} quantum Monte Carlo\cite{tanaka1995,zhang2018,mitra2015,saritas2018} and coupled cluster methods.\cite{Gao2020,takaki2021,Hait2019} Each method comes with its own strengths
and weaknesses, regimes of applicability, and computational cost and scaling.  For example, among the more computationally inexpensive of such techniques is density functional theory (DFT), a popular electronic structure method that predicts material properties from first principles. However, it is well-documented that approximate DFT struggles to accurately describe} the strongly correlated electron systems for which the TMO class is prototypical.\cite{schron_energetic_2010,cococcioni_energetics_2019,zhang_explanation_2014,Cohen792,mori-sanchez_many-electron_2006,isaacs_prediction_2020,georges_strong_2013} By strongly correlated, in this context, we refer to the poor description of the electronic structure associated with localized subspaces within the local (LDA) and semi-local (GGA) mean-field approximations to the XC energy functional.\cite{Oregan2011} This results from the challenge of describing exchange and correlation effects with only a local or semi-local functional dependence, and based on a single-determinant reference system.\cite{mori-sanchez_many-electron_2006,Cohen792,Mori-Sanchez2008,cohen2007,Ruzsinszky2007}
 
Chief among the characteristic errors in the available computationally tractable local or semi-local approximate functionals~\cite{schron_energetic_2010,cococcioni_energetics_2019,zhang_explanation_2014}
is the self-interaction error (SIE), that is the spurious exposure of an electron to its own Coulombic potential, and more broadly, the many-body SIE or delocalization error, a tendency to artificially diffuse the electron density.\cite{Cohen792,mori-sanchez_many-electron_2006,isaacs_prediction_2020,georges_strong_2013,perdew1981} Less discussed, but no less pervasive, is the static-correlation error (SCE) that particularly arises in degenerate and multi-valence systems and, more generally, in systems of significant multi-reference character.

%%%%%%%%%%%%%%% MOs of TMOs diagram %%%%%%%%%%%%%%%%%%%%%%%%%%
\begin{figure*}[t]
    \includegraphics[width=1.0\textwidth]{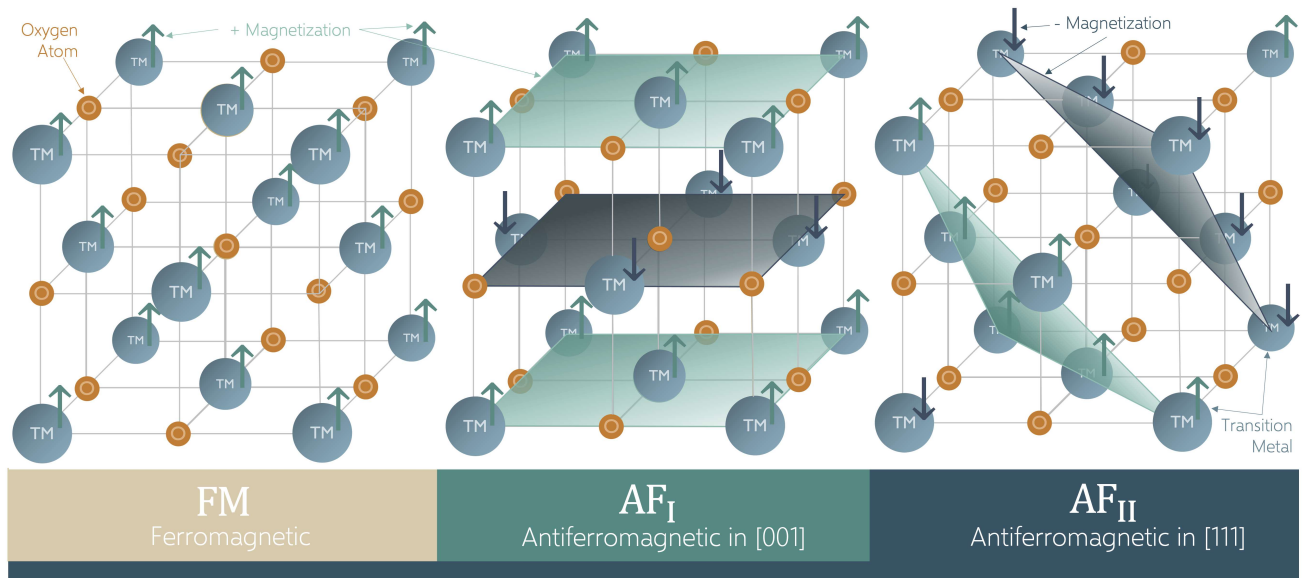}
    \caption{The magnetic orderings of monoxide antiferromagnetic TMOs crystallizing in the rock salt structure, oxygen atoms (\edit{orange}) and TMs (blue). Isomagnetic planes represented by \edit{color gradients}: spin up (\edit{green}) and spin down (navy).}
    \label{mos_of_tmos}
\end{figure*}
%%%%%%%%%%%%%%%%%%%%%%%%%%%%%%%%%%%%%%%%%%%%%%%%%%%%%

Computationally inexpensive techniques that may be interepreted as rectifying SIE, specifically, include the Hubbard-like corrective functionals. Originally inspired by the Hubbard model, DFT+U and related functionals (e.g., DFT+U, DFT+(U-\editt{J$_\textrm{z}$})) append to the approximate DFT Hamiltonian certain compensating energy terms that penalize fractional electron charges within defined subspaces. Moreover, a steadily increasing body of literature finds associations between SCE and the \editt{parameter J$_\textrm{z}$} (\editt{simply denoted J in the literature; renamed here only} to avoid confusion with the Heisenberg $J$).\cite{orhan_first-principles_2020,Lambert2023,georges_strong_2013,Burgess2022} The efficacy of  Hubbard corrective techniques depends on  the underlying exchange correlation functional, as well as the appropriate prescription of the accompanying constants, the Hubbard U and \editt{intra-atomic exchange} coupling \editt{J$_\textrm{z}$} (collectively referred to herein as the Hubbard parameters).\cite{cococcioni_linear_2005,bennett_systematic_2019,kim_quantification_2021,yu_communication_2014,orhan_first-principles_2020}

Many readily-available DFT programs have an implementation of DFT+U in different guises, however
relatively few among them have utilities for calculating  the Hubbard parameters from first principles. In preparation for the present work, we  analysed in detail the numerical performance of a linear-response implementation for calculating the Hubbard U that is embedded in the Projector Augmented Wave (PAW)\cite{blochl_projector_1994} functionality of the open-source plane-wave DFT package  \textsc{Abinit}.\cite{gonze_abinit_2009,Amadon2008} Furthermore, we extended its capability to also calculate the \editt{parameter J$_\textrm{z}$} from first principles. \edit{Our technical work here has led to the development of what is now called the \textsc{Abinit} LRUJ utility.\cite{LRUJtutorial}}

Within the context of Hubbard-corrected DFT, much scientific attention is directed towards understanding the effect of the corrective functionals and their parameters upon bond distances, band gap predictions,\cite{anisimov_band_1991,anisimov_density_1991,anisimov_density_1993} total energy differences,\cite{Wang2021,Yu2020,Dorado2009,Dorado2010,Tompsett2012,Patrick2016,Gopal2017} and other parameterized spectroscopic properties. Meanwhile, comparatively few studies have probed how DFT+U(+J) and related functionals impact metrics of magnetism, such as the magnetic exchange parameters in the Heisenberg model.\cite{Gopal2017,LeBacq2005} When calculated using total energy differences, these parameters become valid ground state DFT quantities, and they thereby become concise indicators of the reliability of DFT+U(+J) total energies. \edit{Magnetic exchange parameters thus constitute expedient indicators of the quality of first-principle DFT+U(+J) total energy differences, which are often supposed to be within the method's regime of reliable applicability, albeit without much concrete literature evidence of that to date. It is essential, and even more so due to its widespread adoption, to work to establish the regime of reliability of first-principles for DFT+U(+J) total-energy differences, and to explore and delineate its remaining weaknesses and ambiguities.}

For these reasons, and given this context, we undertook a multi-pronged investigation designed to systematize the complexities of the magnetic Heisenberg model, first-principles Hubbard corrective functionals, intra-atomic spin parametrization schema, and the PAW method for the prototypical antiferromagnetic TMO, nickel oxide (NiO).

Following a working description of our methodology in Section \ref{Methodology}, we examine, in Section \ref{PAW_Projectors_Results}, the sensitivity of the in situ calculated Hubbard \editt{parameters U and J$_\textrm{z}$}  with respect to the available charge population analyses of the PAW method. 

With Hubbard parameters defined, we map selected points in the energy-magnetization landscape of bulk NiO onto the classical Heisenberg model by calculating its electronic structure in its three main stable, meta-stable or enforceable magnetic configurations (AF$_\textrm{I}$—antiferromagnetic in the [001] direction, AF$_\textrm{II}$—antiferromagnetic in the [111] direction, and ferromagnetic (FM); see Figure \ref{mos_of_tmos}). We extract Heisenberg (inter-atomic) exchange coupling parameters in terms of strict density-functional accessible total energy differences between these magnetic configurations.

In Sections \ref{Results_intra_spin_moment} and \ref{Results_interatomic_spin}, we explore the potential of incorporating not only total energies but also local moments in the mapping onto the Heisenberg
model, and we assess several different ways
to perform this. In doing so, we effectively reevaluate Heisenberg exchange coupling calculation conventions and scrutinize a variety of moment estimation techniques, both classical and semi-classical. Adopting this framework enables a direct and rigorous assessment of the efficacy of 6+ corrective Hubbard functionals via the quality of the Heisenberg exchange coupling parameters as compared to both experiment (i.e., magnon frequencies or inelastic neutron scattering) and ostensibly more comprehensive theory (e.g., hybrid functionals). We relay this comparative analysis in Sections \ref{Hubbard Functionals}-\ref{Hubbard Functionals Explain}, finishing with a closer look at the effect of Hubbard treatment on the O-$2p$ sites in Section \ref{Oxygen}.

\subsection{\label{nickel_oxide}NiO}

As a first-row transition metal harboring several accessible oxidation states, nickel is a relevant candidate for many sustainable technologies. Its monoxide, NiO, is promising as a p-type semiconductor with multiple technological applications, such as in  photocathodes,\cite{materna2020,WAHYUONO2021119507,hsu2012} lithium-ion battery cathodes for  vehicles,\cite{xia2022} OH- carrier battery cathodes for portable electronics,\cite{osti2020} efficiency enhancers in solar cells,\cite{Irwin2008} among others. It is 
relatively non-toxic, 
has a wide and tunable band gap (around 3.5-4.0 eV),\cite{YANG2008,mi2021,Irwin2008,Hugel1983}  and is relative stable. 

NiO crystallizes above the Néel temperature (523 K) in the rocksalt structure (space group \textit{Fm3m}) with an experimental lattice parameter consistently estimated at around 4.17 \AA $\ (7.88\ a_0$) across a variety of diffraction studies. \cite{cracknell_space_1969,roth_multispin_1958,roth1958,eto2000,balagurov2013,balagurov2016,smith1936,peck2012,cheetham1983} Other experiments indicate that Ni sites in bulk NiO take on magnetic moments between 1.96 $\mu_B$\cite{balagurov2016} and 2.2 $\mu_B$.\cite{fernandez1998,neubeck1999,cheetham1983,brok2015} While under certain conditions the TMO can be observed in many magnetic orderings, its ground state magnetic ordering (MO) is AF$_\textrm{II}$, the isomagnetic planes of which alternate antiferromagnetically in the [111] direction (see schematic in Figure \ref{mos_of_tmos}). \cite{cracknell_space_1969,roth1958,roth_multispin_1958}
NiO AF$_\textrm{I}$ and FM magnetic states yield experimental lattice parameters of 4.168 and 4.171 \AA, respectively.\cite{shimomura1956}

Bulk NiO is a well-studied system with a wealth of literature attesting to its total energy difference parameters. Its status as the prototypical Mott-Hubbard insulator renders NiO a good benchmark system through which to observe the effect of SIE corrective techniques on total energies. To date, relatively little is known of the systemic effects of the Hubbard parameters on magnetic exchange interactions in NiO, specifically of the Heisenberg model.

\section{\label{Methodology}Methodology}

\subsection{\label{DFTUFormalism}The Formalism of DFT+U\protect}

Inspired by the Hubbard model,\cite{Hubbard1963} DFT+U\cite{anisimov_band_1991,anisimov_density_1991,dudarev1998,anisimov_density_1993,himmetoglu2014} offers effective treatment of SIE in well-localized electronic states while minimising additional computational expense.\cite{oregan_linear_2012} These corrective functionals are applied to pre-defined localized subspaces that are supposed to be poorly described by the underlying XC functional.\cite{Oregan2011} The total energy in DFT+U is defined as
\begin{equation}\label{LDAplusU}
E_{\textrm{DFT+U}}[\hat{\rho} 
]=E_{\textrm{DFT}}[\hat{\rho}
]+E_{\textrm{U}}[\{n_{mm'}^{i\sigma}\}],
\end{equation}
where \edit{$\hat{\rho}$ is the Kohn-Sham  density
matrix, $E_{\textrm{DFT}}[\hat{\rho}]$} is the approximate DFT total energy, $E_{\textrm{U}}$ is the corrective term, and $\{n_{mm'}^{i\sigma}\}$ refers to the set of projected Kohn-Sham density matrices of the local atomic subspaces that we wish to treat \edit{for SIE on atom $i$ with spin $\sigma$ (e.g., those spanned by atomic-like $3d$ orbitals on Ni atoms). Although all electronic subspaces are afflicted with SIE to some extent, not all are treated with Hubbard corrections, only those with some occupancy matrix eigenvalues   significantly deviating from integer values and, correspondingly, spectral weight adjacent to the band-gap edges. Strictly speaking, many-body SIE is uniquely defined at present only for open quantum systems  in their totality, and partitioning by subspaces implies that pragmatic choices are being made.}

The original derivations of DFT+U considered
the energy, $E_{\textrm{Hub}}$, associated with the Hubbard
model and a single-determinant assumption for the 
wave-function,  together with a
double counting term, $E_{\textrm{dc}}$, for
such a component of the latter that may be already 
included in the underlying Hartree and XC energy.
The double-counting correction is usually 
considered to be dependent solely on $N^{i\sigma}$, the total electron occupation of each subspaces 
and spin component. The exact form of $E_{\textrm{dc}}$ is not precisely known, a consequence of the fact that the XC constituent of the Kohn-Sham Hamiltonian is mutually unknown. One rotationally invariant formulation of DFT+U is known as the Dudarev functional,\cite{dudarev1998,Anisimov_1997}
\begin{align}\label{sriform}
E_{\textrm{U}}[\edit{\{n_{mm'}^{i\sigma}\}}]&=E_{\textrm{Hub}}[\{n_{mm'}^{i\sigma}\}]-E_{\textrm{dc}}[\{N^{i\sigma}\}]\\ \nonumber
&= \sum_{i,\sigma} \frac{\textrm{U}^i_\textrm{eff}}{2} \textrm{Tr}[\textbf{n}^{i\sigma} (1-\textbf{n}^{i\sigma})].
\end{align}

This formulation of DFT+U makes it obvious that the corrective terms preceded by the Hubbard parameters \edit{penalize fractional occupation of atomic orbitals, for positive $\textrm{U}^i_\textrm{eff}$ values.} When an eigenvalue of $\textbf{n}^{i\sigma}$ assumes an integer value, the corresponding contribution to  $E_\textrm{U}$ is zero. By contrast, when that eigenvalue  is maximally fractionalized at 0.5, the contribution to $E_\textrm{U}$ is also maximized.

In Eq. (\ref{sriform}), the effective parameter is U$_{\textrm{eff}}=\textrm{U}-\textrm{J}_\textrm{z}$, where U is the Hubbard parameter and \editt{J$_\textrm{z}$} is the \editt{intra-atomic exchange}  parameter, \editt{more often denoted J but renamed here merely to avoid confusion
with inter-atomic Heisenberg parameters}. The degree of corrective power of the functional, then, is determined by the magnitudes of U and \editt{J$_\textrm{z}$}. In the literature, U$_{\textrm{eff}}$ and U are often referenced interchangeably depending on the consideration given to \editt{J$_\textrm{z}$}.\cite{Kulik2015} An alternative and more elaborate functional that we reference here as DFT+U+\editt{J$_\textrm{z}$}, in which \editt{J$_\textrm{z}$} is  not simply a mitigator of U, is described in Liechtenstein \textit{et al}.\cite{Lichtenstein1995}
For atomic \textit{s}-orbitals, the Liechtenstein functional reduces to the Dudarev functional of Eq. (\ref{sriform}) with U$^i=\textrm{U}_{\textrm{eff}}$. 

\editt{It is worth noting that the J$_\textrm{z}$ parameter computed and employed here pertains, as defined, only to the spin population present in collinear spin DFT. It is elsewhere sometimes termed the Hund's coupling strength because it is affiliated with Hund's coupling-related exchange splitting that manifests in collinear-spin DFT. Strictly speaking, however, to measure that 
mechanism's strength, one would look instead at the energy pre-factor of the effective interaction of the form $\mathbf{S}\cdot\mathbf{S}$. This for vector spin $\mathbf{S}$, either on a per-spatial-orbital or per-atomic-subspace basis, depending on the preferred definition as opposed to its z-axis projection $S_z$.} It is \editt{further} worth noting that, based on recent investigations into the suitability of DFT+U-like functionals in correcting SIE and SCE, \editt{the two functionals used here, when} applied to \textit{s}-orbitals including terms proportional to a positive-valued first-principles J$_\textrm{z}$, do not succeed in correcting for SCE \editt{and may even increase the magnitude of SCE}.\cite{Burgess2022}

Which Hubbard functional, and which subspaces require numerical attention in excess of the underlying XC functional, are both topical inquiries that this article does not directly address for NiO. For the purposes of brevity, we restrict our assessment to DFT+U, DFT+U$_\textrm{eff}$ and DFT+U+\editt{J$_\textrm{z}$}, built upon the Perdew-Burke-Ernzerhof (PBE) GGA\cite{PBE1996} XC functional. We first test the effect of these functionals when in situ treatment is administered to the Ni \textit{3d} orbitals alone (requiring calculation and application of U$^d$ and \editt{J$^d_\textrm{z}$}). Then, we test the same of functionals when in situ treatment is  administered to both the Ni \textit{3d} and O \textit{2p} orbitals together (requiring calculation and application of U$^d$, \editt{J$^d_\textrm{z}$}, U$^p$ and \editt{J$^p_\textrm{z}$}; collectively referred to in each functional as U$^{d,p}$ and \editt{J$^{d,p}_\textrm{z}$}). More information is provided in Section \ref{CompDetails}.

\subsection{\label{SCFLinResp}Calculation of the Hubbard Parameters via Linear Response\protect}

The \editt{Hubbard parameters U and J$_\textrm{z}$} are intrinsic, measurable ground state properties of any multi-atomic system treated with a given approximate XC functional and thus objects that admit  first-principles calculation.\cite{kulik2006,himmetoglu2014} They are related to the interaction contribution to spurious curvature of the total energy with respect to subspace occupation and magnetization, respectively. When properly calculated, the Hubbard parameters are expected to appropriately set the strength of the corrective functionals and precisely cancel the systematic errors characteristic of approximate DFT.\cite{kulik2006,himmetoglu2014,Pickett1998}

\edit{Many methodologies for the in situ calculation of the Hubbard U parameter exist, such as Minimum Tracking Linear Response,\cite{moynihan2017,Linscott2018} the constrained Random Phase Approximation (cRPA),\cite{Aryasetiawan2004,Aryasetiawan2006} and the more recent methodology based on density functional perturbation theory (DFPT),\cite{Timrov2018} to name a few. We focus here on the method from Cococcioni and de Gironcoli,\cite{cococcioni_linear_2005} who, following from the work of Pickett \textit{et al.},\cite{Pickett1998} developed a practical protocol for calculating U from first-principles called the Self-Consistent Field (SCF) Linear Response method. This method} eliminates reliance on external empirical data and paves the road for more accurate simulations of obscure or yet-to-be discovered materials. We refer the reader to these references for a more comprehensive picture of the theory and the mathematical formalism behind the SCF linear response.\cite{Pickett1998,cococcioni_linear_2005,cococcioni_thesis,wu2006,oregan2016}

The SCF linear response approach begins with the application of a small, uniform perturbation to the external potential of the subspace for which U is under assessment. The change in electronic occupation induced by that perturbation is then monitored. The occupancy response is, for small perturbations, ordinarily expected to be a linear function of the perturbation's magnitude.

Formulated practicably, the SCF linear response procedure for calculating U is as follows.\cite{Linscott2018,Lambert2023} To begin, we converge a calculation to the ground state. We then individually (in parallel restarts) apply several \edit{weak perturbations of strength $\{\alpha\}$ in equal magnitude to the up and down spin channels of the external potential.
In the basis of the orbitals that define the subspace $i$, this contribution to the external potential, $V^{i,\sigma}_{m m'}$, is defined in terms of the subspace
projection operator $P^{i}_{m m'}$ by 
\begin{equation}\label{alphaPert}
    V^{i,\sigma}_{m m'}=\alpha P^{i}_{m m'} \delta_{m' m}.
\end{equation}
Here, $P^{i}_{m m'}$ is replaced   
by $\tilde{P}^{i}_{m m'}$ via Eq. (\ref{psuedo_projection_operator}) in the PAW case, which may be approximated in different ways as we later discuss.}
Once the potential perturbation is applied, but before the density is calculated and the Hamiltonian updated to start the new iteration, the spin-up and spin-down charge occupations on the perturbed atom change in response.

Next, we must distinguish between the interacting response and the response arising from a conjectured system in which electrons do not interact (i.e., one with Kohn-Sham (KS) energy).\cite{dederichs1984,cococcioni_thesis,himmetoglu2014} Within this system, we're interested only in the change in total non-interacting occupation ($n^\uparrow_0+n^\downarrow_0$) of subspace $\mathbb{S}$ with respect to $\alpha$, which we call the non-interacting response function, $\chi_0$, defined as
\begin{align}\label{chi0}
\chi_0=\frac{d(n^\uparrow_0+n^\downarrow_0)}{d\alpha}.
\end{align}
%
%%%%%%%%%%%%%%% Linear Response Plots  %%%%%%%%%%%%%%%%%%%%%%%%%%%%
\begin{figure}[b]
    \centering
    \includegraphics[width=1.02\linewidth]{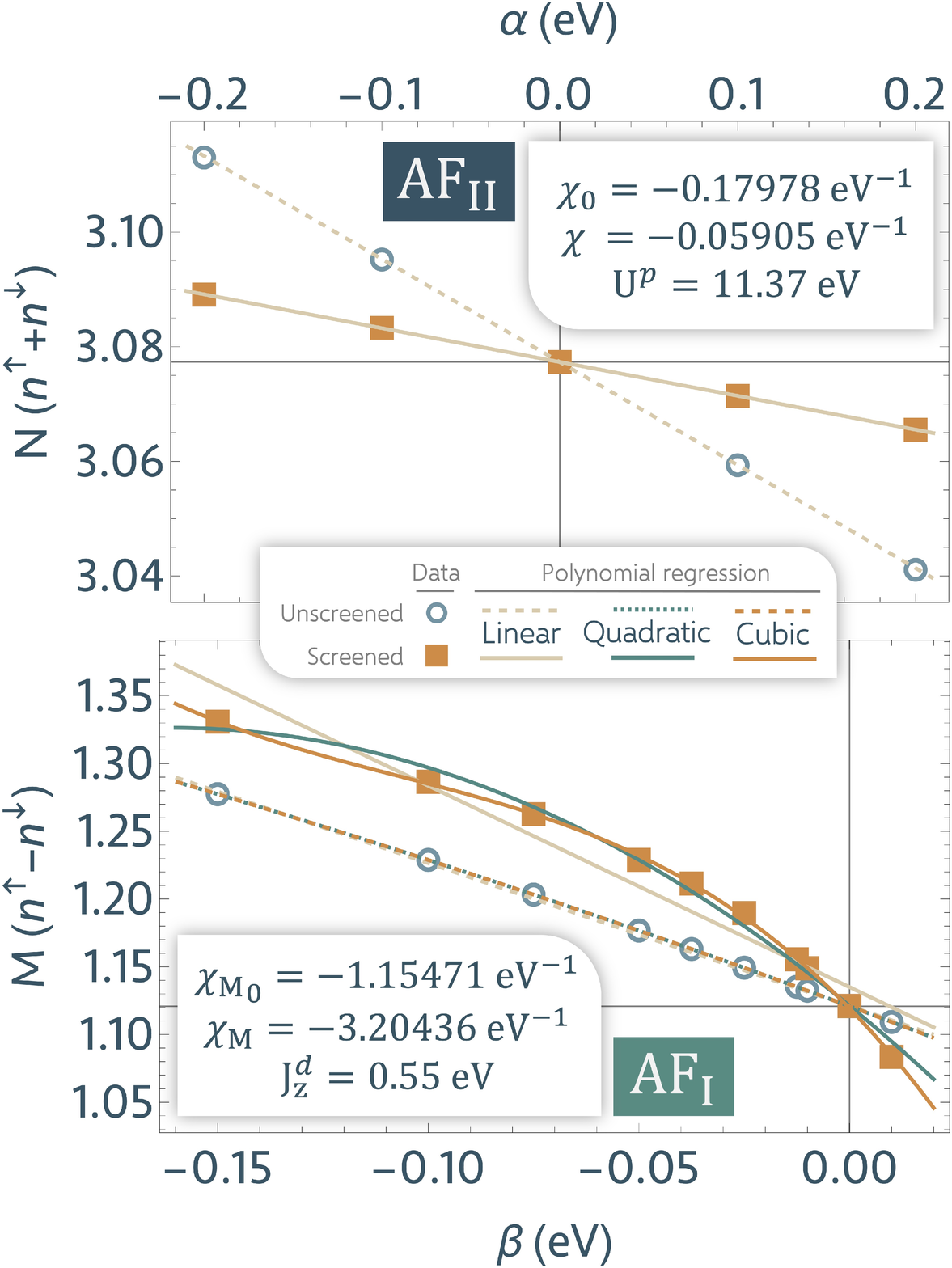}
    \caption{(Top) Example of the linear response (LR) procedure with first-order polynomial regression only, conducted to find the Hubbard \emph{U} for \emph{O} \textit{2p} orbitals on \emph{AF$_\textrm{II}$} \emph{NiO}. (Bottom) \editt{\emph{J$_\textrm{z}$} parameter} LR for \emph{Ni} \textit{3d} orbitals on \emph{AF$_\textrm{I}$} \emph{NiO}. Line color indicates first, second or third order polynomial fit on the data. Axes are aligned with ground state (unperturbed) data. Note: $\chi_0$,$\chi_{\textrm{M}_0}$ data points and fits are transformed to account for technical considerations.\cite{diemix_elaborate}}
    \label{LinearResponsePlot}
\end{figure}
%%%%%%%%%%%%%%%%%%%%%%%%%%%%%%%%%%%%%%%%%%%%%%%%%%%%%%%

We harvest $n^\uparrow_0$ and $n^\downarrow_0$, the non-interacting spin-up and spin-down occupations, respectively, after the first self-consistent iteration. The system then attempts to screen the perturbation, effectively compensating for the disturbance by reorganising its charge once again until, after the last self-consistent iteration, it reaches an equilibrium state. The derivative of this equilibrium occupation with respect to the perturbation magnitude is the interacting response function,
\begin{align}\label{chi}
\chi=\frac{d(n^\uparrow+n^\downarrow)}{d\alpha},
\end{align}
where $n^\uparrow$ and $n^\downarrow$ are the ground state occupations of $\mathbb{S}$ harvested after the last self-consistent iteration.
Finally, we invert the response functions and take their difference to acquire the Hubbard U via the Dyson equation
\begin{align}\label{Dyson2}
U=\chi_0^{-1}-\chi^{-1}.
\end{align}

The \edit{linearity condition}  manifests primarily in the nature of $\chi$ and $\chi_0$ as functions. However, the response behavior is not always linear. Figure \ref{LinearResponsePlot} demonstrates this for an exceptionally ill-behaved system. With some exceptions, linearity is expected in the limit of small perturbations, but this region does not always exist, particularly when the response is excessively hard (e.g., close to full filling). If the perturbations are too large, of course one can generally expect some non-linear behavior. Non-linearity, and sometimes asymmetry across the zero-perturbation axis, is also expected if the system has a particularly shallow energy landscape, (that is, when the response is excessively soft, e.g., close to a phase transition). To compensate for such anomalies, we may fit a regression function (typically polynomial) to the data and differentiate it at $\alpha=\beta=0.0$ eV to find the response functions.

Ideally, U is calculated for a cell of infinite size such that the perturbed subspace is isolated from its periodic images. Since this is unfeasible computationally, U must be converged with respect to an increasing number of atoms, organized into a roughly cubic supercell to isotropically distribute the effect of the perturbation.

The \editt{intra-atomic exchange  parameter J$_\textrm{z}$} can be calculated similarly,\cite{Lambert2023,Linscott2018}
\edit{following the proposal of Ref. \citen{Himmetoglu2011}.}
However, instead of monitoring the change in total subspace occupancy as a function of the applied perturbation $\alpha$, the \editt{J$_\textrm{z}$} relates to changes in magnetization, or difference between the up and down spin occupancies (i.e., $n^\uparrow-n^\downarrow$), in response to perturbations \edit{$\{\beta\}$} applied in positive magnitude to the spin-up potential and in negative magnitude to the spin-down potential, \edit{specifically (adding PAW augmentation as required)
\begin{align}\label{betaPert}
    V^{i,\uparrow}_{m m'}=\beta P^{i}_{m m'} \delta_{m' m}, \quad
    V^{i,\downarrow}_{m m'}=-\beta P^{i}_{m m'} \delta_{m' m}.
\end{align}
}
Noting the minus-sign convention, within SCF linear response \editt{J$_\textrm{z}$} is given by\cite{Lambert2023}
\begin{align}\label{Jtechnical}
\textrm{J}_\textrm{z}&=\frac{d\beta}{d\left(n^\uparrow-n^\downarrow\right)}-\frac{d\beta}{d(n_0^\uparrow-n_0^\downarrow)} \nonumber \\
&=\chi_\textrm{M}^{-1}-\chi_{\textrm{M}_0}^{-1}.
\end{align}

In Appendix \ref{Chi_0_Equivalency} we show that $\chi_0$ and $\chi_{\textrm{M}_0}$ are equivalent, a particularly \edit{convenient result for verifying a correct implementation. We note in passing that Eqs. (\ref{Dyson2}) and (\ref{Jtechnical}) have been derived for use within approximate collinear-spin Kohn-Sham DFT, where only the $z$-projection of spin magnetic moments is considered in the treatment of exchange-correlation effects. The extension of such error quantities to approximate non-collinear spin DFT is a worthy topic outside the scope of this investigation.}

\editt{We emphasize that we denote the intra-atomic exchange parameter $\textrm{J}_\textrm{z}$ as such only to avoid confusion with the inter-atomic Heisenberg exchange parameters $J$, not because it is thought to differ significantly from the familiar J in the context of DFT+U type functionals. We note that, defined as they are here, the $U$ and \editt{J$_\textrm{z}$} can be thought of as error descriptors for SIE and SCE, respectively, in the approximate exchange-correlation functional within the subspace-bath decoupling picture of DFT+U rather than as absolute interaction strengths.  Furthermore, on the topic of interpretation and terminology, we note that \editt{J$_\textrm{z}$} is an \emph{exchange} parameter only 
in the magnetic sense of that word, and it may comprise a significant or even dominant contribution from \emph{correlation} (as distinct from exchange, in the sense of accessibility at the Hartree-Fock level of theory). Indeed, as calculated here it will, by construction, contain a `spin-flip exchange' contribution, which is a contribution to correlation as it does not appear in a Hartree-Fock treatment.~\cite{Bernardi_2020}.}

Based on these definitions, notwithstanding, one may deduce that changes in the atomic magnetic moments---quantities that are informed by the MO of the system---will induce noticeable differences in the Hubbard parameters. That is, the U and \editt{J$_\textrm{z}$} for a particular subspace in FM NiO is expected to be different than those calculated for AF$_\textrm{I}$ or AF$_\textrm{II}$ NiO. However, it is not clear whether the total energies resulting from the application of these MO-specific parameters to their corresponding system remain comparable to each other. We test and discuss this point in Section \ref{Hubbard Functionals}.

\subsection{\label{PAW_Pop_Analysis}PAW-based population analysis\protect}

The special ingredient in both the linear response determination of the Hubbard parameters as well as their host corrective functionals is the subspace occupation matrix $\textbf{n}^{i\sigma}$ and its elements $\{n_{mm'}^{i\sigma}\}$. The DFT+U($\pm$\editt{J$_\textrm{z}$}) occupation matrix elements are defined as
\begin{equation}\label{nocc}
n_{mm'}^{i\sigma}=\sum_{k,v} f_{kv}^\sigma \langle \psi_{kv}^\sigma | \phi_{m'}^i \rangle \langle \phi_{m}^i | \psi_{kv}^\sigma \rangle,
\end{equation}
where $\psi_{kv}^\sigma$ is the Kohn-Sham (KS) orbital corresponding to k-point $k$ and band index $v$, $\phi_{m}^i$ is the localized basis function assigned to magnetic quantum number $m$ on atom $i$, and $f_{kv}^\sigma$ is the Fermi-Dirac distribution function applied to the KS orbitals.

The machinery that isolates charged subspaces for population analysis are the projector functions. The occupation matrix in Eq. (\ref{nocc}) is a generalized projection of the KS density matrix onto a localized basis set of choice, where that choice is typically limited by the 
options made available by the developers of the DFT+U($\pm$\editt{J$_\textrm{z}$}) implementation 
at hand. 
Reliance on projector functions that are unsuitable reflections of the correct localization, either by excessive localization, unphysical discontinuities, or by excessive overlap,\cite{glenn_thesis} will result in erroneous population analysis and, by consequence, inefficient or otherwise defective corrective protocol.\cite{cococcioni_linear_2005}

In \textsc{Abinit}, the DFT+U($\pm$\editt{J$_\textrm{z}$}) formalism is built into the program's Projector Augmented Wave (PAW)\cite{blochl_projector_1994} functionality, so the population analysis relies heavily on the PAW projectors. While an understanding of PAW and its pseudopotential generation schemata is not imperative for the following sections, we encourage the reader to consult, for a more global understanding of the PAW functionality, Refs.~\onlinecite{blochl_projector_1994,kresse_ultrasoft_1999,holzwarth_updated_2019,holzwarth_projector_2001,torrent_electronic_2010,Amadon2008} It suffices to know that the namesake PAW projectors\cite{blochl_projector_1994} are one of three predefined basis sets fed into DFT algorithms via the pseudopotential apparatus. These projectors are designed to isolate a spherically symmetric, element-specific region---an augmentation sphere---about each atomic site, using a relatively small cutoff radius $r_c$.

Within PAW, the DFT+U($\pm$\editt{J$_\textrm{z}$}) occupation matrix elements are obtained via the pseudo (PS) wavefunctions ${\widetilde{\Psi}}_{kv}^{\sigma}$ by computing
\begin{equation}\label{PAW_nocc}
    n_{mm'}^{i\sigma}=\sum_{k,v}{f_{kv}^\sigma \langle {\widetilde{\Psi}}_{kv}^{\sigma} | {\widetilde{P}}_{mm'}^i | {\widetilde{\Psi}}_{kv}^{\sigma} \rangle}.
\end{equation}

Assuming that the charge of the subspace under scrutiny is localized about the atom, we may use the definition of a local PAW PS operator (Eq. (11) in Reference \citen{blochl_projector_1994}) to expand the pseudized projection operator ${\widetilde{P}}_{mm'}^i$ in terms of the KS projection operator $P_{mm'}^i$ and the PAW basis functions: the all-electron (AE) basis $\phi_{\sss I}$, the PS basis $\widetilde{\phi}_{\sss I}$, and the projectors $\widetilde{p}_{\sss I}$. That is
\begin{align}\label{psuedo_projection_operator}
{\widetilde{P}}_{mm'}^i =P_{mm'}^i {}&+ \sum_{IJ}{| {\widetilde{p}}_{\sss I}\rangle\langle \phi_{\sss I}|P_{mm'}^i| \phi_{\sss J}\rangle\langle{\widetilde{p}}_{\sss J}|} \\ \nonumber
 {}&-\sum_{IJ}{|{\widetilde{p}}_{\sss I}\rangle\langle {\widetilde{\phi}}_{\sss I}|P_{mm'}^i| {\widetilde{\phi}}_{\sss J}\rangle\langle{\widetilde{p}}_{\sss J}|}, 
\end{align}
where $I$ is a shorthand index assigned to sets of ($i,\ell_i,m_{\ell_i},n_{i}$), which respectively refer to the site index, angular quantum number, magnetic quantum number, and PAW partial-wave index. Similarly, an analogous set of numbers ($j,\ell_j,m_{\ell_j},n_{j}$) pertains to a different site $J$. When Eq. (\ref{psuedo_projection_operator}) is inserted into Eq. (\ref{PAW_nocc}), three terms emerge, two of which cancel because a requisite in the formation of the PAW basis functions assumes that $\sum_{{\sss I}}|{\widetilde{p}}_{\sss I}\rangle\langle{\widetilde{\phi}}_{\sss I}|=1$ is maintained inside the augmentation region. Acknowledging that $\phi_{\sss I}=\phi_{l_i,m_i{,n}_i}=\phi_{n_i}Y_{l_i,m_i}$, we are left with
\begin{equation}
    n_{mm'}^{i\sigma}=\sum_{IJ}{\rho_{{\sss IJ}}^\sigma\langle\phi_{n_i}|P_{mm'}^i|\phi_{n_j}\rangle},
\end{equation}
where $\phi_n$ are the radial parts of the PAW AE basis functions, and the PAW-sphere density matrix is
\begin{equation}
    \rho_{{\sss IJ}}^\sigma=\sum_{k,v}{f_{kv}^\sigma \langle {\widetilde{\Psi}}_{kv}^{\sigma} |{\widetilde{p}}_{\sss I}\rangle\langle{\widetilde{p}}_{\sss J}| {\widetilde{\Psi}}_{kv}^{\sigma} \rangle}.
\end{equation}

The distribution of charge density about an atom is, in practice, oblivious to the sharp truncation of these PAW projectors at the cutoff radius; even the charge of supposedly localized orbitals often escapes beyond the confines of the augmentation sphere. Care must be taken in selecting an appropriate projector, as the selection can have far-reaching numerical implications.\cite{geneste_dftu_2017,bengone_implementation_2000,torrent_electronic_2010,holzwarth_updated_2019,holzwarth_projector_2001,kresse_ultrasoft_1999}

Four PAW occupation matrix projector options are available in \textsc{Abinit}. For simplicity and ease of reference, we refer to each option by the integer value of its \textsc{Abinit} input file variable, \texttt{dmatpuopt} (Density MATrix for PAW+U OPTion), which may take on values one through four.\cite{dmatpuopt_doc} When \texttt{dmatpuopt=1}, occupations are projections on atomic orbitals,\cite{Amadon2008} per the definition
\begin{equation}\label{dmat1}
    n_{m,m\prime}^{i,\sigma}=\sum_{IJ}{\rho_{{\sss IJ}}^\sigma\left\langle\phi_{n_i}\middle|\phi_{0}\right\rangle\left\langle\phi_{0}\middle|\phi_{n_j}\right\rangle}, 
\end{equation}
where $\phi_{0}$ are normalized, bound state atomic eigenfunctions pertaining to PAW partial wave index $n_i=1$, drawn from the list of AE wavefunctions pertaining to the subspace in question.
\footnote{PAW datasets generated via the \textit{atompaw} code\cite{holzwarth_projector_2001} will always feature a normalized, bound atomic eigenfunction as the first atomic wavefunction of a PAW dataset. Step 4 on Page 2 of Ref. \cite{holzwarth_notes_nodate} states clearly that \textit{atompaw} mandates the use of ``atomic eigenfunctions related to valence electrons (bound states)'' as the partial waves included in the PAW basis. Therefore, all PAW datasets generated by \textit{atompaw}, including the JTH sets on PseudoDojo,\cite{Jollet2014} have atomic eigenfunctions as the first atomic wavefunctions of the correlated subspace. Their normalization, however, depends on the pseudo partial wave generation scheme. \textit{atompaw} provides two options for this scheme: the Vanderbilt or the Blöchl. We are able to reasonably infer %, based on a few key statements in Torrent and Holzwarth's user guide,\cite{holzwarth_notes_nodate} and an \href{https://forum.abinit.org/viewtopic.php?f=9&t=3335&p=10247&hilit=dmatpuopt\#p10247}{\textsc{Abinit} forum response} from Amadon in 2016, 
that the JTH table of PAW datasets,\cite{Jollet2014} available on the PseudoDojo website, match all criteria as a suitable dataset with which one may use \texttt{dmatpuopt=1}.}
Importantly, the integrals $\langle\phi_{n_i}|\phi_{0}\rangle$ and $\langle\phi_{0}|\phi_{n_j}\rangle$ are computed within the PAW augmentation sphere.

By contrast, when \texttt{dmatpuopt=2}, which is the default setting, occupations are integrated in PAW augmentation spheres of charge densities decomposed by angular momenta, per
\begin{equation}\label{dmat2}
    n_{m,m\prime}^{i,\sigma}=\sum_{IJ}{\rho_{{\sss IJ}}^\sigma\left\langle\phi_{n_i}\middle|\phi_{n_j}\right\rangle}. 
\end{equation}

Again, $\left\langle\phi_{n_i}\middle|\phi_{n_j}\right\rangle$ is an integral defined only in the PAW augmentation region.\cite{Amadon2008} Eq. (\ref{dmat2}) may be derived from Eq. (\ref{PAW_nocc}) by setting\cite{bengone_implementation_2000,wang_local_2016}
\begin{align}\label{dmat2_projector}
    P_{mm'}^i(\textbf{r},\textbf{r}') =& 1_\circ(\textbf{r})\delta(|\textbf{r}'-\textbf{R}_\textbf{i}|-|\textbf{r}-\textbf{R}_\textbf{i}|) \\ &\times {Y}_{\ell m}(\hat{\textbf{r}})Y_{\ell m'}^\ast({\hat{\textbf{r}}}'), 
    \nonumber
\end{align}
where $\delta$ is the Dirac-Delta function that effectively counts spatial overlap, $1_\circ\left(\textbf{r}\right)$ is a step function equal to unity when $\textbf{r}$ is inside the augmentation region and zero elsewhere, and ${Y}_{\ell m}$ are the spherical harmonics.

When \texttt{dmatpuopt $\geq$ 3}, occupations take the same form as Eq. (\ref{dmat1}) scaled by normalization constants. When \texttt{dmatpuopt=3} or \texttt{dmatpuopt=4},
\begin{equation}\label{dmat34}
    n_{m,m\prime}^{i,\sigma}=\frac{1}{\mathcal{N}_{0}^{(\texttt{dmatpuopt}-2)}}\sum_{IJ}{\rho_{{\sss IJ}}^\sigma\langle\phi_{n_i}|\phi_{0}\rangle\langle\phi_{0}|\phi_{n_j}\rangle},
\end{equation}
where $\mathcal{N}_0$ is a normalization constant representing the overlap between atomic wavefunctions of the bounded AE basis function inside the PAW augmentation sphere, delimited by cutoff radius $r_c$, per
\begin{equation}\label{dmat_norm}
    \mathcal{N}_0=\int_0^{r_c}{\phi_{0}^2(r)\ dr}.
\end{equation}

The use of \texttt{dmatpuopt=3} is equivalent to Scheme $o_B$ in Eq. (C2) in Geneste \textit{et al},\cite{geneste_dftu_2017} wherein the PAW-truncated atomic orbitals $\phi_{0}$ are effectively renormalized according to the percentage of the total orbital charge inhabiting the augmentation region. When \texttt{dmatpuopt=4}, both the $\phi_{0}$ and the part of the KS orbital falling inside the augmentation sphere are supposedly renormalized (only if the KS orbital matches the atomic orbital in shape), so that $\mathcal{N}_0$ is squared in the denominator.

Based on the description in Timrov \textit{et al.},\cite{timrov_self-consistent_2021}, the plane wave DFT code \textsc{Quantum} ESPRESSO\cite{Giannozzi_2009,Giannozzi_2017} uses none of these options (see Eqs. (9), (10) and (A4) in Timrov \textit{et al.}). Instead, this schema takes the full PS atomic orbitals and augments them with an overlap operator defined in terms of the overlap of the PAW basis functions. In this way, the population analysis avoids relying on orbitals truncated at the PAW cutoff radius and is conducted without significant approximation. For this reason, we calculate the Hubbard U via linear response first in \textsc{Quantum} ESPRESSO for use as a benchmark to determine which PAW-based population analysis involving truncated orbitals is most effective. The results of this investigation are itemized in Table \ref{dmatpuopt_options}.

%%%%%%%%%%%%%%% Superexchange Schematic %%%%%%%%%%%%%%%%%%%%%%%%%%%
\begin{figure*}
    \centering
    \includegraphics[width=1.0\linewidth]{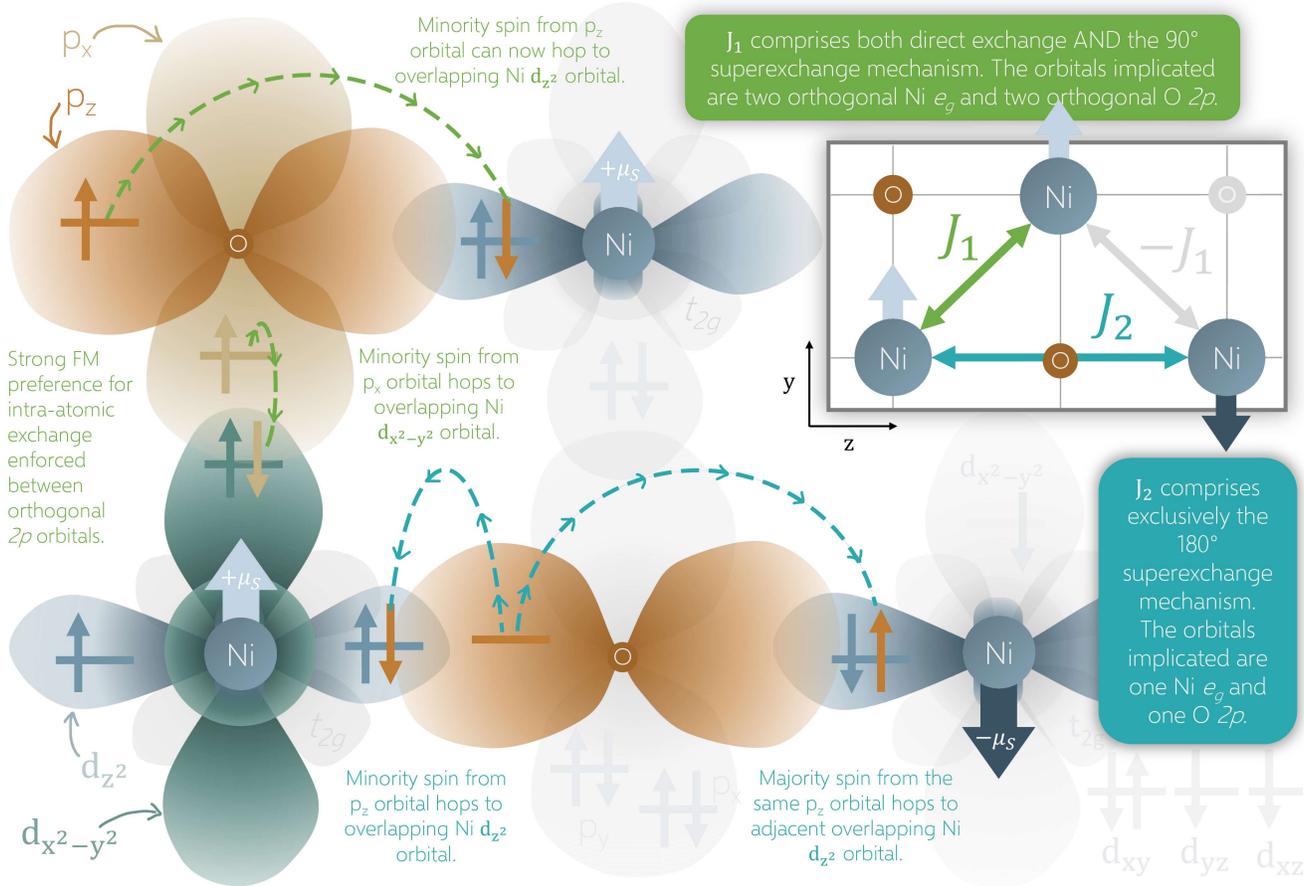}
    \caption{Schematic of xy-plane (z=0) of an \emph{AF$_\textrm{II}$ NiO} in the rocksalt structure, demonstrating how the crystal field splitting influences the preferred orientation of the Heisenberg interatomic exchange interactions J$_1$ (green) and J$_2$ (teal) corresponding, respectively, to the interactions between nearest neighbor transition metal sites and next-nearest neighbor TM sites. More information is provided in Section \ref{HeisenbergModel}.}
    \label{Superexchange_Schematic}
\end{figure*}
%%%%%%%%%%%%%%%%%%%%%%%%%%%%%%%%%%%%%%%%%%%%%%%%%%%%%%%%

\subsection{\label{HeisenbergModel}The Heisenberg Model\protect}

The prevalence of DFT calculations in predicting properties of quantum materials has impeled the development of several effective spin Hamiltonian methods that are compatible with electronic structure theory. Many of these classical methods rely on the justifiable assumption that electron spin, and thus electron magnetic moments, can be treated as discrete vectors, which maintain rigidity in their magnitudes despite rotation.

The simplest effective spin Hamiltonian is the classical Heisenberg model, \edit{which long pre-dates DFT and} assumes that the magnetic moments of electrons are localized, thus defining the Hamiltonian
\begin{align}\label{HeisHam}
H=-\sum_{ij}J_{ij} \textbf{S}_i\cdot\textbf{S}_j, 
\end{align}
where \textit{i} and \textit{j} represent all magnetic centers of the system under consideration, and $\textbf{S}_{i}$ and $\textbf{S}_{j}$ are their corresponding total atomic spin vectors. In the literature there is no unique formulation of Eq. (\ref{HeisHam}), and variation exists based on the definition of the exchange coupling parameters $J_{ij}$ and the constants (i.e., -1, 2, $1/2$) that they absorb. Table \ref{Heisenberg_Conventions} outlines these variations in convention and how they relate to that used here.

We dissect Eq. (\ref{HeisHam}) in the following sections to further deduce how it may best  operate in a DFT context.

%%%%%%%%%%%%%%% Hamiltonian Conventions %%%%%%%%%%%%%%%%%%%%%%%%%%
\begin{table}
    \arrayrulecolor{black}
    \centering
    \begin{tabular}{c|c}
    \hline
    Convention & $J^*_{ij}$ \\ \hline
    \rule{0pt}{4ex} {\color{copper} $\displaystyle H= - \sum_{ij}J^*_{ij}\ \textbf{S}_i\cdot\textbf{S}_j$} & {\color{copper} $J_{ij}$} \\ \hline
    \rule{0pt}{4ex} $\displaystyle H=\gamma \sum_{ij}J^*_{ij}\ \textbf{S}_i\cdot\textbf{S}_j$ & $\frac{-1}{\gamma} J_{ij}$ \\
    \rule{0pt}{4ex} $\displaystyle H=\gamma \sum_{\langle ij \rangle}J^*_{ij}\ \textbf{S}_i\cdot\textbf{S}_j$ & $\frac{-2}{\gamma}J_{ij}$ \\
    \rule{0pt}{4ex} $\displaystyle H=\gamma \sum_{i>j}J^*_{ij}\ \textbf{S}_i\cdot\textbf{S}_j$ & $\frac{-2}{\gamma}J_{ij}$ \\
    \end{tabular}
    \caption{Exchange coupling parameters $J^*_{ij}$ across various conventions of the Heisenberg Hamiltonian in terms of the  exchange coupling parameters $J_{ij}$ used here, where $J^*_{ij}$ are the parameters and $\gamma$ the real constant pertaining to that particular convention. The Heisenberg convention used here (orange) assumes that there is no interaction between one magnetic site and itself. Sometimes this assumption is made explicit by replacing the sum over $ij$ with that over $i\neq j$. The use of angle brackets denotes a set of all pairs $\{i,j\}$ excluding $\{j,i\}$. A similar table may be found in He \textit{et al.}\cite{he2021}}
    \label{Heisenberg_Conventions}
\end{table}
%%%%%%%%%%%%%%%%%%%%%%%%%%%%%%%%%%%%%%%%%%%%%%%%%%%%%%

\subsubsection{\label{ExchangeParameters}Exchange Coupling Parameters\protect}

The Heisenberg model quantifies the phenomenon of interatomic exchange, an effective long-range manifestation of the Pauli exclusion principle, which forbids doubly occupied quantum states. Exchange in the intra-atomic short-range, by contrast, leads to Hund's rules and is several orders of magnitude stronger than its interatomic counterpart.\cite{Skomski2008} The exchange coupling parameters between magnetic atom centers, $J_{ij}$, can be considered spin-based measures of repulsion between electrons on neighboring atoms. When $J_{ij}>0$, a ferromagnetic interaction is preferred, meaning the atoms will adopt the same spin orientation, while $J_{ij}<0$ indicates the atomic spins would rather oppose each other in an antiferromagnetic ordering.

We consider two exchange parameters (refer to the inset in Figure \ref{Superexchange_Schematic} for visualization) following Refs. \citen{archer_exchange_2011,logemann2017,zhang_pressure_2006}. The first is exchange between nearest neighbor (NN) Ni atoms ($J_1$), wherein the overlap of the $t_{2g}$ states of the $3d$ orbitals on NN Ni atoms gives rise to a strongly AF direct exchange. This antiferromagnetic disposition is mitigated, however, by a primarily FM $90^\circ$-oriented indirect exchange, also known as superexchange. This round-the-corner exchange mechanism is strongly mediated by the intra-atomic Coulomb exchange between orthogonal $2p$ orbitals on the pivotal oxygen atom. The schematic in Figure \ref{Superexchange_Schematic} illustrates why we expect this superexchange to favor ferromagnetic alignment. Overall, we expect the FM superexchange to overpower the AF direct exchange, meaning that $J_1$ will adopt a positive value overall.

The schematic in Figure \ref{Superexchange_Schematic} also illustrates why we expect $J_2$, a quantifier of next-nearest neighbor (NNN) exchange interactions separated by $180^\circ$, to adopt a profoundly negative value, indicating strong preference for AF alignment.

With these phenomena in mind, we map the electronic Hamiltonian onto the Heisenberg model and look at $J_1$ and $J_2$ of NiO. We opt to compute the exchange coupling parameters using total energy differences per Ni-O pair resulting from relaxing different magnetic configurations of the same NiO solid in the rocksalt structure. The parameters may be computed as
\begin{align}\label{J1}
    J_1&=\frac{1}{16}(E_{\textrm{AF}_\textrm{I}}-E_\textrm{FM}) \textrm{, and} \\ \label{J2}
    J_2&=\frac{1}{48}(4E_{\textrm{AF}_\textrm{II}}-E_\textrm{FM}-3E_{\textrm{AF}_\textrm{I}}),
\end{align}
where the $E_\textrm{MO}$ are the total energies per Ni-O pair achieved by relaxing NiO in each of its three main stable or meta-stable magnetic configurations. These orderings are depicted in Figure \ref{mos_of_tmos}. A similar figure can be found in Ref. \citen{NiOmagnet}.

\subsubsection{\label{SpinVectorTreatment}Spin Parametrization and Magnetic Moment\protect}

An electron with quantum numbers ($n$,$\ell$,$m_{\ell}$,$m_s$) will have spin magnetic moment $\boldsymbol{\mu_s}=-g_e\mu_B \boldsymbol{s}$, where $g_e \approx 2.002318$ is the g-factor for a free electron and $\mu_B=\frac{e\hbar}{2m}=\frac{1}{2}$ [a.u.] is the Bohr magneton.\cite{resnickBook1976,coey2010magnetism} The total spin magnetic moment is the sum of spin magnetic moments of all electrons on that atom $\boldsymbol{\mu_\textbf{S}}=-g_e \mu_B \textbf{S}$ (where $\textbf{S}$ is the summation over electrons).\cite{coey2010magnetism,Li2021} Referencing atomic magnetic moments in units of $\mu_B$, then, $\textbf{S}\ [\mu_B]=-\boldsymbol{\mu_\textbf{S}}/g_e$. In atomic units, $\textbf{S}$ [a.u.] $=-2\boldsymbol{\mu_\textbf{S}}/g_e$.

Conventionally, the $\textbf{S}_{i}$ of Eq. (\ref{HeisHam}) are considered (a) projections $(\textrm{S}_z)_{i}$ of the quantum spin vector $\textbf{S}_{i}$ on the quantization axis, and (b) of unit magnitude in accordance with nickel's expected oxidation state of +2, corresponding to an atomic magnetic moment of $+2\ \mu_B$ ($=1\ $ in atomic units, hence $\boldsymbol{\mu_S}^2\approx 1$). We seek to reevaluate these conventions given the multitude of DFT-accessible population analyses that can influence, to a large degree, the predicted magnetic moment on Ni atoms.

For example, in Section \ref{PAW_Pop_Analysis}, we found that the PAW population analysis extends naturally to a definition of the magnetic moment that is both atom-centered and restricted to the augmentation sphere. That is, for magnetic centers, we take
\begin{align}\label{Sz_moment_AS}
(\textrm{S}^{\textrm{AS}}_z)_i
=\int_{|\textbf{r}|=0}^{r_c} (\rho^\uparrow(\textbf{R}_i+\textbf{r})-\rho^\downarrow(\textbf{R}_i+\textbf{r}))\ d\textbf{r} .
\end{align}

However, because the PAW augmentation spheres do not envelop all the charge in the system, leaving some interstitial charge uncounted, this option is expected to produce moments smaller than expected. We may address this by normalizing the subspace occupations via the projector, as in Section \ref{PAW_Pop_Analysis}, and calculating
\begin{align}\label{Sz_moment_PAW}
(\textrm{S}^\textrm{PAW}_z)_{i}=\sum_{m} n^{i,\uparrow}_{m,m} - n^{i,\downarrow}_{m,m}, 
\end{align}
which, by consequence, encodes the choice of PAW projectors described in Section \ref{PAW_Pop_Analysis}. Alternatively, we may obtain magnetic moments via the extraction of $ \textrm{S}_z $ from the entire spin density (SD), by using
\begin{align}\label{Sz_moment_SD}
(\textrm{S}^\textrm{SD}_z)_{i}=\frac{1}{N_\textrm{Ni}}\int |\rho^\uparrow(\textbf{r})-\rho^\downarrow(\textbf{r}) |\ \ d\textbf{r},
\end{align}
where $N_\textrm{Ni}$ is the number of Ni atoms in the system. While it produces larger magnetic moments, this method is accompanied by the particular limitation that the atom-specific magnetic moments are no longer resolved, and any moments formed near the oxygen atoms are absorbed into those of the magnetic centers. It is conventional in the relevant literature to consider oxygen atoms as mere extensions or mediators of the magnetic character of the TM magnetic centers, thereby restricting assessment of exchange interactions to those between Ni atoms. A similar study conducted by Logemann \textit{et al.}\cite{logemann2017} investigates the appropriateness of this convention. In using Eq. (\ref{Sz_moment_SD}), we follow this study.

It should be noted, however, that the Heisenberg Hamiltonian in Eq. (\ref{HeisHam}) calls for the use of the quantum spin vectors $\textbf{S}_{i}$, as opposed to their scalar and DFT-accessible counterparts, the expectation values of the \textit{z}-axis projection of the spin angular momentum $ \textrm{S}_z $. Suppose we instead consider an approximate quantum model in which $|\textbf{S}|=\sqrt{\langle \textbf{S}^2 \rangle}$. Coupling the spins on each magnetic site such that $s=1$ for the two unpaired spins on Ni together, the expectation value of the spin angular momentum of an atomic system is
\begin{align}\label{S_moment_quantum}
|\textbf{S}|&=\sqrt{\langle \textbf{S}^2 \rangle}= \hbar \sqrt{s(s+1)}\textrm{S}_z  \\ \nonumber
&=\sqrt{2}\  \textrm{S}_z \ \textrm{[a.u.]}.
\end{align}

That is, in a sort of crude quantum approximation, we scale the DFT-accessible magnetic moments, which may be thought of as time-averages of precessing, canted spins, by a factor of $\sqrt{2}$. To our knowledge, this approximation has been made to magnetic properties in other contexts---such as in the works of Harrison,\cite{harrison_heisenberg_2007} Kotani \textit{et al},\cite{kotani2008} and Wan \textit{et al}\cite{wan_calculations_2006}---but not tested systematically against other DFT-accessible spin objects for NiO.

Once we settle on an appropriate treatment of the spin vector, we can examine how it factors into the numerical calculation of the exchange coupling parameters. Eqs. (\ref{J1}) and (\ref{J2}) can be considered simplifications of a more generalized set of total energy difference equations under the Heisenberg model, derived by considering all NN and NNN of one Ni atom (or Ni-O pair) in the cell. They are solutions to a system of linear equations mapping the total energy of each MO onto the Heisenberg model, solutions that simplify when we impose assumptions that the spin magnetic moments of all atoms are a) integers and b) the same regardless of the cell's magnetic environment. That is, the magnitudes of the atomic magnetic moments  are $\vert\textrm{S}_\textrm{FM}\vert=\vert\textrm{S}_{\textrm{AF}_{\textrm{I}}}\vert=\vert\textrm{S}_{\textrm{AF}_{\textrm{II}}}\vert=1$ in
\begin{align}\label{totalenergyheis}
    &E_\textrm{FM}=E_0-12 J_1 \textrm{S}_\textrm{FM}^2-6 J_2 \textrm{S}_\textrm{FM}^2 \\ \nonumber
    &E_{\textrm{AF}_{\textrm{I}}}=E_0+4 J_1 \textrm{S}_{\textrm{AF}_{\textrm{I}}}^2-6 J_2 \textrm{S}_{\textrm{AF}_{\textrm{I}}}^2 \\ \nonumber
    &E_{\textrm{AF}_{\textrm{II}}}=E_0+6 J_2 \textrm{S}_{\textrm{AF}_{\textrm{II}}}^2\ .
\end{align}

Suppose we assume instead that $\vert\textrm{S}_\textrm{FM}\vert\neq \vert\textrm{S}_{\textrm{AF}_\textrm{I}}\vert\neq \vert\textrm{S}_{\textrm{AF}_\textrm{II}}\vert\neq1$. Since DFT uses the non-integer spin degree of freedom to deepen the total energy well of each MO, and since that relaxed magnetization is accessible, we can perhaps explicitly use the final magnetization to better weight the energetic contributions of each MO to the exchange coefficients. Then the system of linear equations in Eqs. (\ref{totalenergyheis}), solved for $J_1$ and $J_2$, becomes
\footnote{Eqs. (\ref{J1wmagn}) and (\ref{J2wmagn}) are true only in atomic units, where $\mu_B=1/2$ and the numerical value of these units is absorbed into the other constants of the equations. Thus, total energies expressed in eV must each be multiplied by a factor of four for Eqs. (\ref{J1wmagn}) and (\ref{J2wmagn}) to hold true.}
\begin{widetext}
\begin{align}\label{J1wmagn}
    J_1 &= \frac{E_{\textrm{AF}_{\textrm{II}}}(\textrm{S}_{\textrm{AF}_{\textrm{I}}}^2-\textrm{S}_\textrm{FM}^2)+E_{\textrm{AF}_{\textrm{I}}}(\textrm{S}_{\textrm{AF}_{\textrm{II}}}^2+\textrm{S}_\textrm{FM}^2)-E_\textrm{FM}(\textrm{S}_{\textrm{AF}_{\textrm{I}}}^2+\textrm{S}_{\textrm{AF}_{\textrm{II}}}^2)}{4\left[3 \textrm{S}_{\textrm{AF}_{\textrm{II}}}^2 \textrm{S}_\textrm{FM}^2+\textrm{S}_{\textrm{AF}_{\textrm{I}}}^2\left(\textrm{S}_{\textrm{AF}_{\textrm{II}}}^2+4 \textrm{S}_\textrm{FM}^2\right)\right]}, \quad \mbox{and}
\\
\label{J2wmagn}
    J_2 &= \frac{ E_{\textrm{AF}_{\textrm{II}}}(\textrm{S}_{\textrm{AF}_{\textrm{I}}}^2+3 \textrm{S}_\textrm{FM}^2)-3 E_{\textrm{AF}_{\textrm{I}}} \textrm{S}_\textrm{FM}^2-E_\textrm{FM} \textrm{S}_{\textrm{AF}_{\textrm{I}}}^2}{6\left[ 3 \textrm{S}_{\textrm{AF}_{\textrm{II}}}^2 \textrm{S}_\textrm{FM}^2+\textrm{S}_{\textrm{AF}_{\textrm{I}}}^2\left(\textrm{S}_{\textrm{AF}_{\textrm{II}}}^2+4 \textrm{S}_\textrm{FM}^2\right)\right]}.
\end{align}
\end{widetext}

Note that if we set $\vert \textrm{S}_\textrm{FM}\vert=\vert\textrm{S}_{\textrm{AF}_{\textrm{I}}}\vert=\vert\textrm{S}_{\textrm{AF}_{\textrm{II}}}\vert=1$ then Eqs. (\ref{J1wmagn}) and (\ref{J2wmagn}) reduce to Eqs. (\ref{J1}) and (\ref{J2}).

Exchange coupling parameters that incorporate differences in magnetization across MOs are not immediately comparable to those reported in previous works. We find this is not so much an artefact of the assumption that $\vert\textrm{S}_\textrm{FM}\vert=\vert\textrm{S}_{\textrm{AF}_{\textrm{I}}}\vert=\vert\textrm{S}_{\textrm{AF}_{\textrm{II}}}\vert=\sigma$ (where $\sigma$ is a constant) so much as it reflects the assumption that $\sigma=1$. As a comparative intermediary, therefore, we look at the exchange interaction parameters when the sum of electrons in a Ni atom corresponds directly to the magnitude of the Ni magnetic moment in the ground state magnetic configuration of NiO (i.e., when $\sigma^2=\textrm{S}_{\textrm{AF}_\textrm{II}}^2$ [a.u.]). Under this approach, Eqs. (\ref{J1wmagn}) and (\ref{J2wmagn}) reduce to
\begin{align}\label{J1wmagnsimple}
    J_1 &= \frac{1}{16\ \textrm{S}_{\textrm{AF}_\textrm{II}}^2}\left( E_{\textrm{AF}_{\textrm{I}}}-E_\textrm{FM} \right), \quad \mbox{and}
\\ \label{J2wmagnsimple}
    J_2 &= \frac{1}{48\ \textrm{S}_{\textrm{AF}_\textrm{II}}^2} \left( 4 E_{\textrm{AF}_{\textrm{II}}}-3 E_{\textrm{AF}_{\textrm{I}}}-E_\textrm{FM} \right).
\end{align}

For the purposes of clarity, we will refer to the first method of calculating exchange interaction parameters, those embodied by Eqs. (\ref{J1}) and (\ref{J2}), as Method A. Similarly, Eqs. (\ref{J1wmagnsimple}) and (\ref{J2wmagnsimple}) are henceforth referenced as Method B, and Eqs. (\ref{J1wmagn}) and (\ref{J2wmagn}) are Method C.

%%%%%%%%%%%%%%%%% NiO Characteristics %%%%%%%%%%%%%%%%%%%%%%%%%%%%
\setlength\arrayrulewidth{1pt}
\setlength{\extrarowheight}{1pt}
\begin{table*}
\centering
\arrayrulecolor{white}
\begin{tabular}{r|c|c|c|c|c|c|}

\rslate[\overhang]
{\cellcolor[HTML]{FFFFFF} } & \multicolumn{3}{c|}{\tcw \spc\spc\spc \textbf{DFT LR Calculations} \spc\spc\spc} & \multicolumn{3}{|c|}{\tcw \spc \textbf{DFT+U($\pm$\editt{J$_\textrm{z}$}) Exchange Calculations} \spc} \\ \hline

\rowcolor[HTML]{FFFFFF}[\overhang]
& \cellcolor{tan}{\tcw\spc\spc\spc\spc\spc\spc\textbf{FM}\spc\spc\spc\spc\spc\spc} &
\seacl{\tcw\spc\spc\spc\spc\spc\spc\textbf{AF$_{\textrm{I}}$}\spc\spc\spc\spc\spc\spc} &
\multicolumn{1}{c|}{\cellcolor{navy}{\tcw\spc\spc\spc\spc\textbf{AF$_{\textrm{II}}$}\spc\spc\spc\spc}} &
\multicolumn{1}{|c|}{\cellcolor{tan}{\tcw\spc\spc\spc\spc\spc\spc\spc\textbf{FM}\spc\spc\spc\spc\spc\spc\spc}} &
\seacl{\tcw\spc\spc\spc\spc\spc\spc\spc\textbf{AF$_{\textrm{I}}$}\spc\spc\spc\spc\spc\spc\spc} & \cellcolor{navy}{\tcw\spc\spc\spc\spc\spc\spc\textbf{AF$_{\textrm{II}}$}\spc\spc\spc\spc\spc\spc} \\ \hline

\rowcolor[HTML]{EEF2F4}[\overhang]
\tbd \textbf{\# Atoms} & \multicolumn{3}{c|}{\tbl 64} & \multicolumn{3}{|c|}{\tbl 4} \\ \hline

\rowcolor[HTML]{EEF2F4}[\overhang]
\tbd \textbf{ Experimental a (\AA)} & 4.171\footnote{Refs. \citen{cracknell_space_1969,roth_multispin_1958,roth1958,eto2000,balagurov2013,balagurov2016,smith1936,peck2012,cheetham1983}} & 4.168\footnote{Ref. \citen{shimomura1956}} & \multicolumn{1}{c|}{4.170$^\textrm{b}$} & \multicolumn{1}{|c|}{4.171$^\textrm{a}$} & 4.168$^\textrm{b}$ & 4.17$^\textrm{b}$ \\ \hline

\rowcolor[HTML]{EEF2F4}[\overhang]
\tbd & \multicolumn{3}{c|}{} & \multicolumn{1}{|c|}{} & & \\
\rowcolor[HTML]{EEF2F4}[\overhang]
\tbd & \multicolumn{3}{c|}{} & \multicolumn{1}{|c|}{} & & \\
\rowcolor[HTML]{EEF2F4}[\overhang]
\multirow{-3}{*}{\tbd \textbf{Lattice Vectors}} & \multicolumn{3}{c|}{\multirow{-3}{*}{\tbl\thead{ $[2,0,0]$ \\ $[0,2,0]$ \\ $[0,0,2]$}}} & \multicolumn{1}{|c|}{\multirow{-3}{*}{\tbl\thead{ $[\nicefrac{1}{2},\nicefrac{1}{2},0]$ \\ $[0\ ,\ 0\ ,\ 1]$ \\ $[\nicefrac{1}{2},-\nicefrac{1}{2},0]$}}} & \multicolumn{1}{c|}{\multirow{-3}{*}{\tbl\thead{ $[\nicefrac{1}{2},\nicefrac{1}{2},0]$ \\ $[0\ ,\ 0\ ,\ 1]$ \\ $[\nicefrac{1}{2},-\nicefrac{1}{2},0]$}}} & \multirow{-3}{*}{\tbl\thead{ $[\nicefrac{1}{2},\nicefrac{1}{2},1]$ \\ $[1,\nicefrac{1}{2},\nicefrac{1}{2}]$ \\ $[\nicefrac{1}{2},1,\nicefrac{1}{2}]$}} \\ \hline

\rowcolor[HTML]{EEF2F4}[\overhang]
\tbd\spc\spc\spc\spc \textbf{Energy Tolerance (Ha)} & \multicolumn{3}{c|}{\tbl $1\times10^{-7}$ } & \multicolumn{3}{|c|}{\tbl $1\times10^{-12}$} \\ \hline

\rowcolor[HTML]{EEF2F4}[\overhang]
\tbd \textbf{k-Point Sampling} & \tbl $3\times3\times3$ & $3\times3\times3$ & \multicolumn{1}{c|}{\tbl $5\times5\times5$} & \multicolumn{1}{|c|}{\tbl $40\times20\times40$} & \tbl $40\times20\times40$ & \tbl $30\times30\times30$ \\ \hline

\rowcolor[HTML]{EEF2F4}[\overhang]
\tbd \textbf{Insulating Character} & metal & metal & \multicolumn{1}{c|}{insulator} & \multicolumn{1}{|c|}{insulator} & insulator & insulator \\ \hline

\rowcolor[HTML]{EEF2F4}[\overhang]
\tbd \textbf{U$^d\ |$ J$^d_\textrm{z}$ (eV)} & 0.0 $|$ 0.0 & 0.0 $|$ 0.0 & \multicolumn{1}{c|}{0.0 $|$ 0.0} & \multicolumn{1}{|c|}{6.91 $|$ 0.18} & 6.74 $|$ 0.55 & 5.58 $|$ 0.47 \\ \hline

\rowcolor[HTML]{EEF2F4}[\overhang]
\tbd \textbf{U$^p\ |$ J$^p_\textrm{z}$ (eV)} & 0.0 $|$ 0.0 & 0.0 $|$ 0.0 & \multicolumn{1}{c|}{0.0 $|$ 0.0} & \multicolumn{1}{|c|}{11.64 $|$ 1.55} & 11.30 $|$ 1.87 & 11.37 $|$ 1.82 \\ \hline

\end{tabular}
\caption{Calculation parameters specific to each magnetic ordering of NiO for both the calculation and application of the Hubbard parameters, respectively.
We include here number of atoms, experimental lattice parameter $a$, lattice vectors, converged energy tolerance, k-Point sampling, insulating character, and Hubbard \editt{parameters \emph{U} and \emph{\editt{J$_\textrm{z}$}}} calculated for the Ni $3d$ and $2p$ subspaces respectively, calculated with the population analysis schema of Eq. (\ref{dmat34}) (see Section \ref{PAW_Pop_Analysis} for a discussion of population analysis schemes).}
\label{RawPBE}
\end{table*}
%%%%%%%%%%%%%%%%%%%%%%%%%%%%%%%%%%%%%%%%%%%%%%%%%%%%%%%

\subsection{\label{CompDetails}Computational Details}

We use different runtime parameters for computing the Hubbard parameters and applying them due to the different simulation cell and attendant Brillouin zone sizes needed for each task. This information, for all parameters that varied across tasks and MO (FM, AF$_\textrm{I}$ and AF$_\textrm{II}$), are laid out in Table \ref{RawPBE}. 

Otherwise, consistent across all calculations, we use the Perdew-Burke-Ernzerhof (PBE) GGA\cite{PBE1996} XC functional and the Jollet-Torrent-Holzwarth PBE PAW pseudopotentials (Version 1.1)\cite{Jollet2014} generated via \textit{atompaw}.\cite{holzwarth_projector_2001} 
\edit{Derived from those pseudopotentials, the PAW augmentation sphere cutoff radius $r_c$ used in, for example, Eqs. (\ref{dmat2_projector}) and (\ref{Sz_moment_AS}), is 1.81 $a_0$ for Ni and 1.41 $a_0$ for O.}
We use the experimental lattice parameters pertaining to each MO of NiO specified in Table \ref{RawPBE}, and for those that have metallic character, we use the Marzari-Vanderbilt smearing scheme\cite{marzari_thesis} with 0.005 Ha broadening. Furthermore, we enforce a cutoff energy of 33.1 Ha (44.1 Ha on the  PAW atom augmentation sphere). In terms of the magnetic moment, we extract $\textrm{S}_\textrm{z}$ in accordance with Eq. (\ref{Sz_moment_SD}) using C2x.\cite{rutter_c2x_2018}

For brevity, we test six Hubbard functionals: DFT+U$^d$, DFT+U$^{d,p}$, DFT+U$^d_\textrm{eff}$, DFT+U$^{d,p}_\textrm{eff}$, DFT+U$^d$+J$^d_\textrm{z}$, DFT+U$^{d,p}$+J$^{d,p}_\textrm{z}$. Furthermore, we seek to understand if total energy differences between MOs are most comparable when calculated using Hubbard parameters that are (a) specific to each MO, or (b) the same across MOs. To test (a), we apply each of the aforementioned Hubbard functionals with MO-specific Hubbard parameters (HPs) to its respective magnetically ordered version of NiO (that is, FM HPs applied to FM NiO via each functional, AF$_\textrm{I}$ HPs to AF$_\textrm{I}$ NiO, and so on). To test (b), we apply the U and \editt{J$_\textrm{z}$} from the ground state AF$_\textrm{II}$ MO to every MO of NiO via each of the aforementioned Hubbard functionals (that is, AF$_\textrm{II}$ HPs applied to FM NiO via each functional, AF$_\textrm{II}$ HPs to AF$_\textrm{I}$ NiO, and so on).

\subsubsection{\label{LR_Compute}Linear Response Calculations}

We conduct \textsc{Abinit} linear response calculations via what is now its intrinsic LRUJ algorithm\cite{LRUJtutorial} to determine the Hubbard \editt{parameters U and \editt{J$_\textrm{z}$}} for both the \textit{3d} Ni and \textit{2p} O orbitals (procedure described in Section \ref{SCFLinResp}). For each U (\editt{J$_\textrm{z}$}) parameter, at least five perturbations between $\alpha=\pm$0.2 eV ($\beta=\pm$0.1 eV) are applied to one of the 64 atoms in the supercells of each MO of NiO. Depending on how well these calculations converged, additional perturbations were applied.

%%%%%%%%%%%%%%%%% dmatpuopt table %%%%%%%%%%%%%%%%%%%%%%%%%%%%
\begin{table*}[t]
\centering
\arrayrulecolor{black}
\begin{tabular}{c|cc|cccV{0.1}ccc|cccV{0.1}ccc}
\multicolumn{1}{c}{} & \multicolumn{1}{c}{} & \multicolumn{1}{c}{} & \multicolumn{6}{c}{Ni Linear Response Parameters} & \multicolumn{6}{c}{O Linear Response Parameters} \\ \cline{4-15}
\multicolumn{1}{c}{} & \multicolumn{1}{c}{} & \multicolumn{1}{c}{} & \multicolumn{3}{cV{0.1}}{U} & \multicolumn{3}{c|}{\editt{J$_\textrm{z}$}} & \multicolumn{3}{cV{0.1}}{U} & \multicolumn{3}{c}{\editt{J$_\textrm{z}$}} \\ \cmidrule(lr){4-6}\cmidrule(lr){7-9}\cmidrule(lr){10-12}\cmidrule(lr){13-15}
\multicolumn{1}{c}{MO} & \multicolumn{2}{c|}{ $n^{t,\sigma}_{m,m'}$} & \spc\spc\spc $\chi_0$ \spc\spc\spc & \spc\spc\spc $\chi$ \spc\spc\spc & \spc\spc\spc U$^d$ \spc\spc\spc & \spc\spc\spc $\chi_{\textrm{M}_0}$ \spc\spc\spc & \spc\spc\spc $\chi_{_J}$ \spc\spc\spc & \spc\spc\spc J$^d_\textrm{z}$ \spc\spc\spc & \spc\spc\spc $\chi_0$ \spc\spc\spc & \spc\spc\spc $\chi$ \spc\spc\spc & \spc\spc\spc U$^p$ \spc\spc\spc & \spc\spc\spc $\chi_{\textrm{M}_0}$ \spc\spc\spc & \spc\spc\spc $\chi_{_J}$ \spc\spc\spc & \spc\spc\spc J$^p_\textrm{z}$ \spc\spc\spc \\ \hline
\rowcolor[HTML]{FFFFFF}[\overhang]
\multirow{5}{*}{\cellcolor{tan}{\color{white}}} &  & 1 &-0.8064 & -0.1024 & \textbf{8.53} & -0.8054 & -0.9829 & \textbf{0.22} & -0.0856 & -0.0264 & \textbf{26.16} & -0.0856 & -0.1216 & \textbf{3.46} \\
\rowcolor[HTML]{FFFFFF}[\overhang]
{\cellcolor{tan}} &  & 2 & -0.9856 & -0.1359 & \textbf{6.34} & -0.9854 & -1.1746 & \textbf{0.16} & -0.1901 & -0.0595 & \textbf{11.55} & -0.1901 & -0.2697 & \textbf{1.55} \\
 \rowcolor[HTML]{FFFFFF}[\overhang]
{\cellcolor{tan}{\color{white}\textbf{FM}}} &  & 3 & -0.9858 & -0.1262 & \textbf{6.91} & -0.9826 & -1.1892 & \textbf{0.18} & -0.1914 & -0.0593 & \textbf{11.64} & -0.1914 & -0.2719 & \textbf{1.55}  \\
 \rowcolor[HTML]{FFFFFF}[\overhang]
{\cellcolor{tan}} & \spc\multirow{-4}{*}{\rotatebox{90}{\color{black} \textsc{Abinit}}} & 4 & -1.2010 & -0.1610 & \textbf{5.38} & -1.2010 & -1.3887 & \textbf{0.11} & -0.4281 & -0.1314 & \textbf{5.28} & -0.4280 & -0.6075 & \textbf{0.69} \\ \clineB{2-15}{0.1}
 \rowcolor[HTML]{FFFFFF}[\overhang]
{\cellcolor{tan}} & \multicolumn{2}{c|}{QE} & -0.5008 & -0.1178 & \textbf{6.49} & {\cellcolor{lightgrey}} & {\cellcolor{lightgrey}} & {\cellcolor{lightgrey}} & -0.1176 & -0.0530 & \textbf{10.36} & {\cellcolor{lightgrey}} & {\cellcolor{lightgrey}} & {\cellcolor{lightgrey}} \\ \hline
\rowcolor[HTML]{FFFFFF}[\overhang]
\multirow{5}{*}{\cellcolor{sea}{\color{white}}} &  & 1 & -0.9219 & -0.1064 & \textbf{8.31} & -0.9449 & -2.5779 & \textbf{0.67} & -0.0923 & -0.0277 & \textbf{25.27} & -0.0927 & -0.1545 & \textbf{4.31} \\
\rowcolor[HTML]{FFFFFF}[\overhang]
{\cellcolor{sea}} & {\tcb} & 2 & -1.1289 & -0.1347 & \textbf{6.54} & -1.1577 & -3.3259 & \textbf{0.56} & -0.2054 & -0.0601 & \textbf{11.52} & -0.2060 & -0.3363 & \textbf{1.88} \\
\rowcolor[HTML]{FFFFFF}[\overhang]
{\cellcolor{sea}{\color{white}\textbf{AF$_\textrm{I}$}}} & {\tcb} & 3 & -1.1259 & -0.1311 & \textbf{6.74} & -1.1547 & -3.2044 & \textbf{0.55} & -0.2068 & -0.0620 & \textbf{11.30} & -0.2073 & -0.3390 & \textbf{1.87} \\
\rowcolor[HTML]{FFFFFF}[\overhang]
{\cellcolor{sea}} & \spc\multirow{-4}{*}{\rotatebox{90}{\color{black} \textsc{Abinit}}} & 4 & -1.3750 & -0.1583 & \textbf{5.59} & -1.4107 & -3.1137 & \textbf{0.39} & -0.4634 & -0.1417 & \textbf{4.90} & -0.4643 & -0.7652 & \textbf{0.85} \\ \clineB{2-15}{0.1}
\rowcolor[HTML]{FFFFFF}[\overhang]
{\cellcolor{sea}} & \multicolumn{2}{c|}{QE} & -0.6299 & -0.1244 & \textbf{6.45} & {\cellcolor{lightgrey}} & {\cellcolor{lightgrey}} & {\cellcolor{lightgrey}} & -0.1144 & -0.0516 & \textbf{10.64} & {\cellcolor{lightgrey}} & {\cellcolor{lightgrey}} & {\cellcolor{lightgrey}} \\ \hline
\rowcolor[HTML]{FFFFFF}[\overhang]
\multirow{5}{*}{\cellcolor{navy}{\color{white}}} &  & 1 & -0.3104 & -0.0998 & \textbf{6.80} & -0.3104 & -0.3870 & \textbf{0.64} & -0.0803 & -0.0264 & \textbf{25.49} & -0.0799 & -0.1181 & \textbf{4.04} \\
\rowcolor[HTML]{FFFFFF}[\overhang]
{\cellcolor{navy}} &  & 2 & -0.3791 & -0.1248 & \textbf{5.37} & -0.3790 & -0.4737 & \textbf{0.53} & -0.1784 & -0.0593 & \textbf{11.27} & -0.1779 & -0.2640 & \textbf{1.83} \\
\rowcolor[HTML]{FFFFFF}[\overhang]
{\cellcolor{navy}{\color{white}\textbf{AF$_\textrm{II}$}}} &  & 3 & -0.3793 & -0.1218 & \textbf{5.58} & -0.3795 & -0.4621 & \textbf{0.47} & -0.1798 & -0.0590 & \textbf{11.37} & -0.1793 & -0.2661 & \textbf{1.82}  \\
\rowcolor[HTML]{FFFFFF}[\overhang]
{\cellcolor{navy}} & \spc\multirow{-4}{*}{\rotatebox{90}{\color{black} \textsc{Abinit}}} & 4 & -0.4644 & -0.1428 & \textbf{4.85} & -0.4634 & -0.5206 & \textbf{0.24} & -0.4024 & -0.1320 & \textbf{5.09} & -0.4017 & -0.5953 & \textbf{0.81} \\ \clineB{2-15}{0.1}
\rowcolor[HTML]{FFFFFF}[\overhang]
{\cellcolor{navy}} & \multicolumn{2}{c|}{QE} & -0.2607 & -0.0998 & \textbf{6.18} & {\cellcolor{lightgrey}} & {\cellcolor{lightgrey}} & {\cellcolor{lightgrey}} & -0.1247 & -0.0537 & \textbf{10.59} & {\cellcolor{lightgrey}} & {\cellcolor{lightgrey}} & {\cellcolor{lightgrey}} \\ \hline
\end{tabular}
\caption{PAW-based occupation determination methods (variations itemized by \emph{\texttt{dmatpuopt}} integer for brevity: \emph{1} refers to Eq. (\ref{dmat1}), \emph{2} refers to Eq. (\ref{dmat2}), and \emph{3} and \emph{4} refer to Eq. (\ref{dmat34})) compared with that of QuantumESPRESSO \emph{(QE)} in terms of the effect induced on Hubbard parameters, \emph{U} and \emph{J$_{\textrm{z}}$}, as well as their bare ($\chi_0$) and screened ($\chi$) response function components. Numbers gathered via linear response procedures implemented on both \emph{Ni} and \emph{O} across all magnetic orderings \emph{(MO)} of the 64-atom \emph{NiO} supercell.}
\label{dmatpuopt_options}
\end{table*}
%%%%%%%%%%%%%%%%%%%%%%%%%%%%%%%%%%%%%%%%%%%%%%%%%%%%%%

As described at the end of Section \ref{PAW_Pop_Analysis}, \textsc{Quantum} ESPRESSO augements the pseudo wavefunctions with core contributions without truncating at the PAW cutoff radius. Therefore, we further undertook \textsc{Quantum} ESPRESSO linear response calculations for the Hubbard U on Ni and O across all magnetic orderings.

For each U parameter calculated with \textsc{Quantum} ESPRESSO, five equally spaced perturbations between $\alpha=\pm$0.2 eV are applied to one of the 64 atoms in the supercells of each MO of NiO. Runtime parameters remain the same as those for \textsc{Abinit} except the following: The energy is relaxed self-consistently to within a threshold tolerance of $10^{-9}$ Ha, and a kinetic energy of 25 Ha for wavefunctions is administered alongside an energy cutoff of 250 Ha for the density. For all MOs, we use Monkhorst-Pack\cite{monkhorst-pack1976} grid k-point sampling of $2 \times 2\times 2$.

\section{\label{Results}Results and Discussion}

\subsection{\label{PAW_Projectors_Results}PAW Projectors and the Hubbard Parameters}

Switching between PAW population analysis schemes induces noticeable numerical differences in the resulting Hubbard parameters for both Ni $3d$ orbitals and O $2p$ orbitals (see Table \ref{dmatpuopt_options}). It is encouraging that the unscreened response functions $\chi_0$ and $\chi_{\textrm{M}_0}$, pertaining to each projector choice, are reasonably alike. This is expected in accordance with the Corollary in Appendix \ref{Chi_0_Equivalency}.

The Hubbard \editt{parameters U and J$_{\textrm{z}}$} resulting from \texttt{dmatpuopt=2} and \texttt{dmatpuopt=3} are consistently similar, whereas \texttt{dmatpuopt=1} yields the largest in magnitude of these parameters and \texttt{dmatpuopt=4} yields the smallest. These phenomena are concordant with our understanding of the make-up of the response matrices. That is, extrapolations of occupancies beyond the PAW radius increase the overall occupancy (magnetization), thereby decreasing the inverse of the response matrices and reducing U (J$_{\textrm{z}}$).

For Ni, which has well-localized valence electrons, the range of Hubbard parameters is not very wide; by our calculations, 90.47\% of the $3d$ atomic wave functions lie inside the PAW augmentation sphere. For O $2p$ orbitals, however, this overlap is significantly smaller, around 66.87\%, and thus the range of achievable Hubbard parameters by various population analysis schemes is significantly larger for O than for Ni. It is important, therefore, in treating O $2p$-orbitals, to select an occupancy matrix calculation scheme that is normalized inside the PAW augmentation sphere.

In terms of numerical similarity with the \textsc{Quantum} ESPRESSO-derived U, only \texttt{dmatpuopt=2} and \texttt{dmatpuopt=3} remain viable candidates, with \texttt{dmatpuopt=2} barely outperforming \texttt{dmatpuopt=3} based on numerical nuances. Not only does \texttt{dmatpuopt=4} systematically underestimate U, from a theoretical standpoint (described in Section \ref{PAW_Pop_Analysis}), using the integral of $\phi_{0}$ to renormalize the AE basis function is an approximation that is not entirely justified. While the first AE partial wave listed in a PAW dataset for a particular treated subspace, which \textsc{Abinit} selects as $\phi_{0}$, is required to be a normalized atomic eigenfunction, the second (or third or fourth, etc.) AE basis functions have no such constraints. There's no guarantee that these functions are localized in the PAW augmentation region to the same extent as $\phi_{0}$. By contrast, based on \texttt{dmatpuopt=1}'s systematic overestimation of U, it is clear that at least some normalization is necessary, especially for the oxygen \textit{2p} states.

For these reasons, we focus our attention on the subtle differences between \texttt{dmatpuopt=2} and \texttt{3}. It is worth mentioning that, outside of convergence difficulties arising from the J$_{\textrm{z}}$ calculations for the FM and AF$_\textrm{I}$ MOs, \texttt{dmatpuopt=3} marginally outperforms \texttt{dmatpuopt=2} in terms of quality of linear response and ease of numerical convergence. The metrics used to evaluate this quality were (a) similarity of unscreened response functions $\chi_0$ and $\chi_{\textrm{M}_0}$; (b) average number of SCF cycles needed to converge perturbative calculations; (c) average regression order (i.e., how often did the response functions demonstrate linearity over higher order polynomial regressions), and; (d) the error on the Hubbard parameter, calculated as follows:
\begin{equation}\label{HP_error_eq}
\sigma_\textrm{U}=\sqrt{\left(\frac{\sigma_{\chi_0}}{{\chi_0}^2}\right)^2+\left(\frac{\sigma_\chi}{\chi^2}\right)^2}
\end{equation}
where $\sigma_{\chi_0}$ and $\sigma_{\chi}$ are the unbiased root mean squared errors on the regression fit. The same error assessment is true for J$_{\textrm{z}}$. For \texttt{dmatpuopt=3}, the average $\sigma_\textrm{U}$($\sigma_{\textrm{J}_\textrm{z}}$) across all HPs is $\bar{\sigma}_\textrm{U}= 1.28\%$ ($\bar{\sigma}_{\textrm{J}_\textrm{z}}= 4.36\%$), which is slightly better than \texttt{dmatpuopt=2}'s $\bar{\sigma}_\textrm{U}= 1.41\%$ ($\bar{\sigma}_{\textrm{J}_\textrm{z}}= 5.20\%$). All in all, \texttt{dmatpuopt=3} outperforms its competitor in three of the four aforementioned metrics ((b), (c) and (d), to be precise).

There is a case to be made in opposition to \texttt{dmatpuopt=2} from a theoretical standpoint. Its projection operator, Eq. (\ref{dmat2_projector}), is defined using a Heaviside step function, which operationally ``counts," so to speak, the overlap between two PAW radial basis functions by brute-force. The ground state eigenfunction basis for \texttt{dmatpuopt=3} is underpinned by the same assumptions and approximations that are used in the definition of the bound PAW AE partial waves. The renormalization of projections of all partial waves with a bounded, truncated AE wavefunction reintroduces a crucial, physics-based safeguard into the definition of the occupancy matrix. For this reason, we select \texttt{dmatpuopt=3} as our preferred PAW population analysis. We report the Hubbard parameters pertaining to this population analyisis scheme as well as their associated errors in Table \ref{Final_HPs+Errors}.

\subsection{\label{Results_intra_spin_moment}Intra-atomic spin moment}

Having fossilized the in situ Hubbard parameters, we must give due consideration to the most appropriate Hubbard functional (among the six tested using the AF$_\textrm{II}$ HPs), intra-atomic spin moment calculation (AS, PAW or SD, permutated across $ \textrm{S}_z $ and $|\textbf{S}|$), and MO-based spin parametrization method (Methods A, B, or C). Table I of the Supplemental Material (SM) provides all of this raw data. To make the process of elimination simpler, however, we'll focus on identifying the most effective combination of spin treatments by averaging across all six Hubbard functionals the percent RMS deviation (\% RMSE) of our results for $J_1$ and $J_2$ from those of one particular experiment (exp), specifically
\begin{equation}\label{percentRMSEJ1J2}
\%\ \textrm{RMSE}=\sqrt{\frac{1}{2}\left(\left(\frac{J_1}{J_1^\textrm{exp}}\right)^2+\left(\frac{J_2}{J_2^\textrm{exp}}\right)^2\right)}.
\end{equation}

We select the inelastic neutron scattering experiment of Hutchings \textit{et al}\cite{Hutchings_1972} as our benchmark for this analysis. After averaging Eq. (\ref{percentRMSEJ1J2}) across the six Hubbard functionals, SM Table I reduces to Table \ref{SpinMomentErrors}, where the darker shades of blue coincide with the degree to which error is minimized.

%%%%%%%%%%%%%%%%% Magn and Js Graphs %%%%%%%%%%%%%%%%%%%%%%%%%
\begin{figure*}
    \centering
    \includegraphics[width=1.0\textwidth]{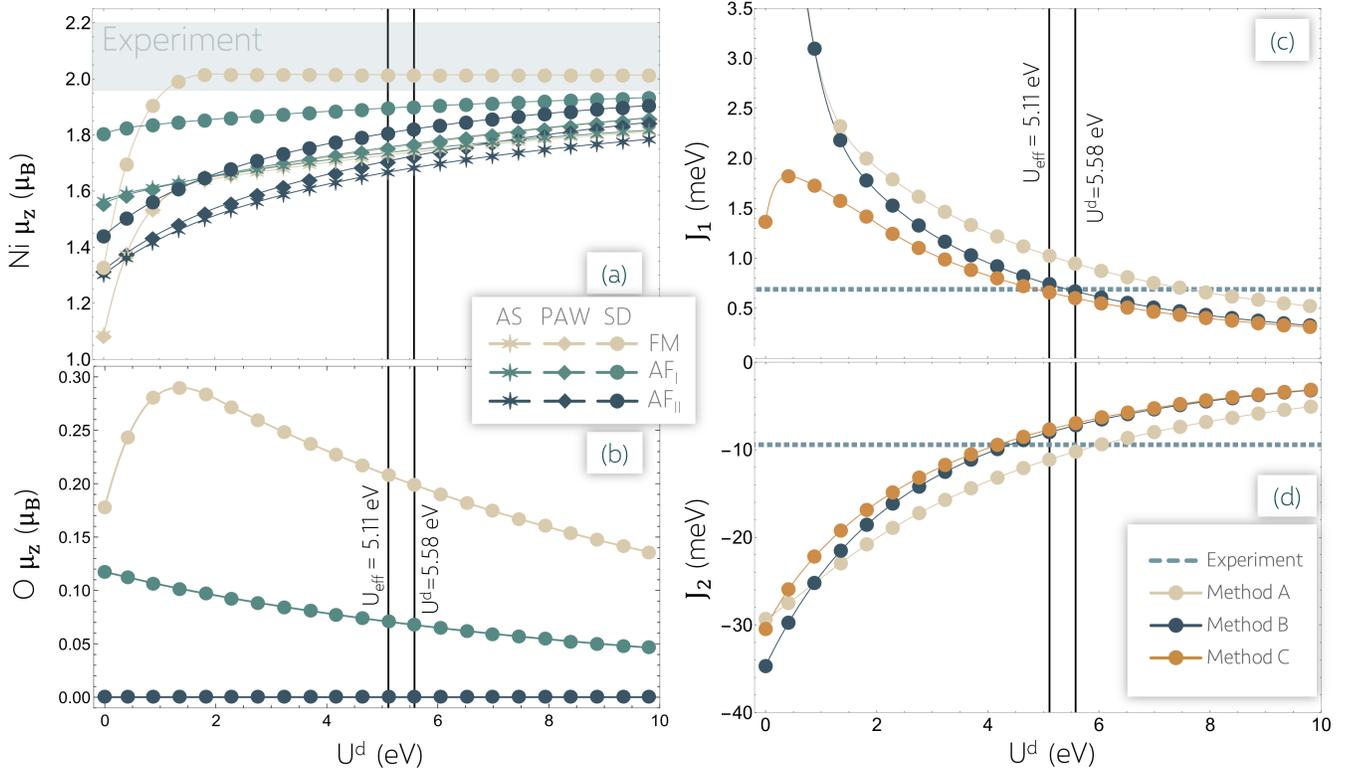}
    \caption{The effect of treating \emph{Ni} \textit{3d} orbitals in the \emph{DFT+U$^d$} functional with varying magnitude of the \emph{AF$_\textrm{II}$} \emph{$\textrm{U}^d_\textrm{eff}=\textrm{U}^d-\textrm{J}^d_\textrm{z}$} (same HPs across all MOs). Plot \emph{(a)} and \emph{(b)} show average nickel and oxygen site magnetic moments $\mu_z$, projected on the z-axis and resolved via magnetic ordering (\emph{FM} for ferromagnetic, \emph{AF$_\textrm{I}$} for antiferromagnetic I and \emph{AF$_\textrm{II}$} for antiferromagnetic II), for each population analysis described in Section \ref{SpinVectorTreatment}; \emph{AS} = augmented sphere (Eq. (\ref{Sz_moment_AS})), \emph{PAW} = PAW occupations (Eq. (\ref{Sz_moment_PAW})), and \emph{SD} = total spin density (Eq. (\ref{Sz_moment_SD})). Plots \emph{(c)} and \emph{(d)} show, respectively, nearest-neighbor and next-nearest neighbor Heisenberg exchange interaction parameters, calculated via different spin-parametrization methods, where intra-atomic spin is calculated as \emph{$\vert \textbf{S}^\textrm{AS}\vert =\sqrt{2}\  \textrm{S}^\textrm{AS}_z $}.}
    \label{NiUvsMagJs}
\end{figure*}
%%%%%%%%%%%%%%%%%%%%%%%%%%%%%%%%%%%%%%%%%%%%%%%%%%%%%

We notice immediately that the \% RMS error is smaller when approximately treating intra-atomic spin as a quantum vector. Compared with its z-axis projection, $| \textbf{S} |$ diminishes differences across population analysis schemes, such that $| \textbf{S}^\textrm{AS} |$ is only subtly favored over other candidates. The preferable MO-based spin parametrization method depends crucially on the use of the quantum vector over the z-axis projection; under the latter treatment, Method A obtains exchange parameters more in line with experiment by a clear margin, indicating a preference for total energy differences with idealized spin-moment values. By contrast, if the magnetic moments are amplified by a factor of $\sqrt{2}$ as with the quantum vector approximation, Methods B and C, which weight the total energies from each MO according to various combinations of their atomic magnetic moments, win out.

%%%%%%%%%%%%% Official Hubbard Parameters %%%%%%%%%%%%%%%%%%%%%%%%%%
\begin{table}[b]
\centering
\arrayrulecolor{black}
\begin{tabular}{|c|c|c|c|}

\cline{2-4}
\rowcolor[HTML]{FFFFFF}[\overhang]
\multicolumn{1}{c|}{(eV)} & {\cellcolor{tan}{\color{white}\textbf{FM}}} & {\cellcolor{sea}{\color{white}\textbf{AF$_\textrm{I}$}}} & {\cellcolor{navy}{\color{white}\textbf{AF$_\textrm{II}$}}} \\ \hline

U$^d$ &\spc\spc\spc $6.91\pm0.13$ \spc\spc\spc&\spc\spc\spc $6.74\pm0.08$ \spc\spc\spc&\spc\spc\spc\spc $5.58\pm0.09$ \spc\spc\spc \\
J$^d_{\textrm{z}}$ & $0.18\pm0.02$ & $0.55\pm0.01$ & $0.47\pm0.02$ \\
U$^p$ & $11.64\pm0.06$ & $11.30\pm0.16$ & $11.37\pm0.12$ \\
J$^p_{\textrm{z}}$ & $1.55\pm0.03$ & $1.87\pm0.04$ & $1.82\pm0.05$ \\
\hline

\end{tabular}
\caption{In situ Hubbard parameters calculated via linear response for \emph{FM}, \emph{AF$_\textrm{I}$} and \emph{AF$_\textrm{II}$} \emph{NiO} and associated errors, calculated via Eq. (\ref{HP_error_eq}), reported in \emph{eV}.}
\label{Final_HPs+Errors}
\end{table}
%%%%%%%%%%%%%%%%%%%%%%%%%%%%%%%%%%%%%%%%%%%%%%%%%%%%%%

To understand this behaviour, we recall that DFT-derived Ni magnetic moments on AF NiO do not typically exceed 2 $\mu_B$ (see Figure \ref{NiUvsMagJs} (a)), meaning that the coefficients by which the AF total energies are weighted exceed unity, thereby amplifying the total energy differences to which the exchange interaction parameters are proportional. Most underlying Hubbard functionals, in and of themselves, generate total energy differences that are too large anyway, so the modifications provided by these magnetic moments are almost always counterproductive in obtaining results that more closely resemble experiment. The only exceptions to this trend occurred when we applied Hubbard functionals using parameters exceeding those calculated via linear response by several
eV. Overestimation of the exchange parameters continues as long as DFT underestimates the magnetic moment of Ni on NiO.

According to Figure \ref{NiUvsMagJs} (a), experimentally consistent AF moments are out of the reach of all population analysis techniques. The additional $\sqrt{2}$ factor accompanying $| \textbf{S} |$ serves to approximately  compensate where DFT falls short, lending itself to overcompensation when coupled with the population analysis techniques that typically predict higher Ni magnetic moments by default (like PAW and SD). For this reason, we see that a relatively underestimated population analysis, like that coming from AS, coupled with the overestimation attributed to the $\sqrt{2}$ factor from $| \textbf{S} |$ can balance out to yield minimized errors. 

%%%%%%%%%%%%%%%% Spin treatment Heat Map %%%%%%%%%%%%%%%%%%%%%%%%%
\begin{table}[b]
\begin{center}
\centering
\begin{tabular}{c|ccc|ccc}
\multicolumn{1}{c}{{\color[HTML]{FFFFFF} }} & \multicolumn{3}{c}{$ \textrm{S}_z  $} & \multicolumn{3}{c}{$|\textbf{S}|$} \\ \cmidrule(lr){2-4}\cmidrule(lr){5-7}
\rowcolor[HTML]{FFFFFF}[\overhang]
\multicolumn{1}{c|}{\small \%RMSE} & \multicolumn{1}{c}{\tcw\cellcolor{tan} \spc\spc A \spc\spc} & \multicolumn{1}{c}{\tcw\cellcolor{navy} \spc\spc B \spc\spc} & \multicolumn{1}{c|}{\tcw\cellcolor{copper} \spc\spc C \spc\spc} & \multicolumn{1}{c}{\tcw\cellcolor{tan} \spc\spc A \spc\spc} & \multicolumn{1}{c}{\tcw\cellcolor{navy} \spc\spc B \spc\spc} & \multicolumn{1}{c}{\tcw\cellcolor{copper} \spc\spc C \spc\spc} \\ \hline
\rowcolor[HTML]{FFFFFF}[\overhang]

AS & {\cellcolor[HTML]{c1cfdb} 50.2} & {\cellcolor[HTML]{ffffff} 105.1} & {\cellcolor[HTML]{edf1f5} 89.8} & {\cellcolor[HTML]{c1cfdb} 50.2} & {\cellcolor[HTML]{9aa7b2} 29.7} & {\cellcolor[HTML]{8c9aa7} 29.4} \\

\rowcolor[HTML]{FFFFFF}[\overhang]
PAW & {\cellcolor[HTML]{c1cfdb} 50.2} & {\cellcolor[HTML]{f5f7f9} 96.3} & {\cellcolor[HTML]{eaeef2} 86.4} & {\cellcolor[HTML]{c1cfdb} 50.2} & {\cellcolor[HTML]{9ba9b4} 30.3} & {\cellcolor[HTML]{9ca9b4} 30.4} \\

\rowcolor[HTML]{FFFFFF}[\overhang]
SD & {\cellcolor[HTML]{c1cfdb} 50.2} & {\cellcolor[HTML]{dfe6ec} 76.9} & {\cellcolor[HTML]{f9fafb} 100.0} & {\cellcolor[HTML]{c1cfdb} 50.2} & {\cellcolor[HTML]{a2afbb} 33.1} & {\cellcolor[HTML]{a9b6c2} 36.3} \\
\end{tabular}
\end{center}
\caption{Percent RMS errors (\emph{\%RMSE}). \emph{AS} spin moments refer to those calculated in atom-centered augmentation spheres (Eq. (\ref{Sz_moment_AS})); \emph{PAW} refers to cumulative differences in PAW occupation matrix elements (Eq. (\ref{Sz_moment_PAW})); \emph{SD} refers to use of the global spin density difference (Eq. (\ref{Sz_moment_SD})). Cell hue represents error magnitude; the darker the blue, the smaller the error.}
\label{SpinMomentErrors}
\end{table}
%%%%%%%%%%%%%%%%%%%%%%%%%%%%%%%%%%%%%%%%%%%%%%%%%%%%%%

This explanation, of course, is not a glowing testimonial for either technique, especially since $ \textrm{S}_z^\textrm{AS} $ performs rather poorly compared to its PAW and SD counterparts. However, the results in and of themselves, independent of their justification, offer concrete, actionable insights to the reader in terms of which spin treatment is best for NiO exchange. Certainly, Table \ref{SpinMomentErrors} offers more organized navigation of a few available DFT-tractable, spin-based objects and their influence on the Heisenberg exchange constants, and the numerical forgiveness awarded by these techniques should not be overlooked. For the purposes of further analysis, therefore, we take Table \ref{SpinMomentErrors} at its face value and both select and recommend $| \textbf{S}^\textrm{AS} |=\sqrt{2}\  \textrm{S}^\textrm{AS}_z $ as the preferred intra-atomic spin moment calculation method. Data pertaining to the $ \textrm{S}_z $ calculations across all population analysis methods is provided in the SM.

%%%%%%%%%%%%%% Hubbard Functional Results %%%%%%%%%%%%%%%%%%%%%%%%%%
\begin{table*}
\begin{center}
\centering
\begin{tabular}{l|cc|cc|cc|l}
\multicolumn{1}{c}{{\color[HTML]{FFFFFF} }} & \multicolumn{2}{c}{\color{tan} Method A} & \multicolumn{2}{c}{\color{navy} Method B} & \multicolumn{2}{c}{\color{copper} Method C} & \\ \cmidrule(lr){2-3}\cmidrule(lr){4-5}\cmidrule(lr){6-7}
\rowcolor[HTML]{FFFFFF}[\overhang]
\multicolumn{1}{c|}{} & \multicolumn{2}{c|}{{\tiny\tcw\cellcolor{tan} \spc $\textrm{S}_\textrm{FM}=\textrm{S}_{\textrm{AF}_{\textrm{I}}}=\textrm{S}_{\textrm{AF}_{\textrm{II}}}=1$} \spc} & \multicolumn{2}{c|}{{\tiny\tcw\cellcolor{navy}\spc $\textrm{S}_\textrm{FM}=\textrm{S}_{\textrm{AF}_{\textrm{I}}}=\mathbf{\textrm{S}_{\textrm{AF}_{\textrm{II}}}}$} \spc} & \multicolumn{2}{c}{{\tiny\tcw\cellcolor{copper}\spc $\textrm{S}_\textrm{FM}\neq\textrm{S}_{\textrm{AF}_{\textrm{I}}}\neq\textrm{S}_{\textrm{AF}_{\textrm{II}}}$} \spc} & \\
\rowcolor[HTML]{FFFFFF}[\overhang]
& {\tcw\cellcolor{tan} $J_1$} & {\tcw\cellcolor{tan} $J_2$} & {\tcw\cellcolor{navy} $J_1$} & {\tcw\cellcolor{navy} $J_2$} & {\tcw\cellcolor{copper} $J_1$} & {\tcw\cellcolor{copper} $J_2$} & \multicolumn{1}{c}{\% RMS Error} \\ 
\rowcolor[HTML]{FFFFFF}[\overhang]
\multicolumn{1}{l|}{Functional} & {\tcw\cellcolor{tan}\spc\spc\spc (meV)\spc\spc\spc\spc} & {\tcw\cellcolor{tan}(meV)} & {\tcw\cellcolor{navy}\spc (meV) \spc\spc} & {\tcw\cellcolor{navy}(meV)} & {\tcw\cellcolor{copper}\spc\spc (meV) \spc\spc\spc} & {\tcw\cellcolor{copper}(meV)} & \\ \cline{1-7}
DFT (PBE) & 8.23 & -29.32 & 9.73 & -34.68 & 1.37 & -30.48 & \multirow{-2}{*}{\includegraphics[width=0.21\textwidth]{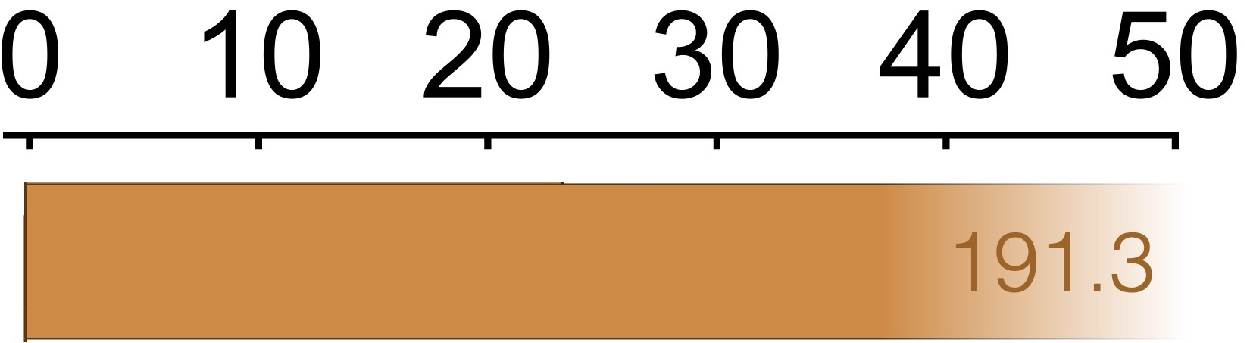}} \\ 
\multicolumn{1}{c}{AF$_\textrm{II}$ NiO Parameters} & & & & & & & \\ \cmidrule(lr){1-1}
DFT + U$^d$ & 0.95 & -10.20 & 0.67 & -7.22 & 0.60 & -6.95 & \\ 
DFT + U$^{d,p}$ & 1.53 & -7.11 & 1.07 & -5.00 & 1.00 & -4.88 & \\ 
DFT + U$^d_\textrm{eff}$ & 1.03 & -11.10 & 0.74 & -8.00 & 0.66 & -7.68 & \\ 
DFT + U$^{d,p}_\textrm{eff}$ & 1.52 & -8.41 & 1.09 & -6.04 & 1.00 & -5.86 & \\
DFT + U$^d$ + J$^d_{\textrm{z}}$ & 0.96 & -10.24 & 0.69 & -7.31 & 0.62 & -7.04 & \\
DFT + U$^{d,p}$ + J$^{d,p}_{\textrm{z}}$ & 1.00 & -7.35 & 0.71 & -5.22 & 0.66 & -5.09 & \multirow{-6}{*}{\includegraphics[width=0.225\textwidth]{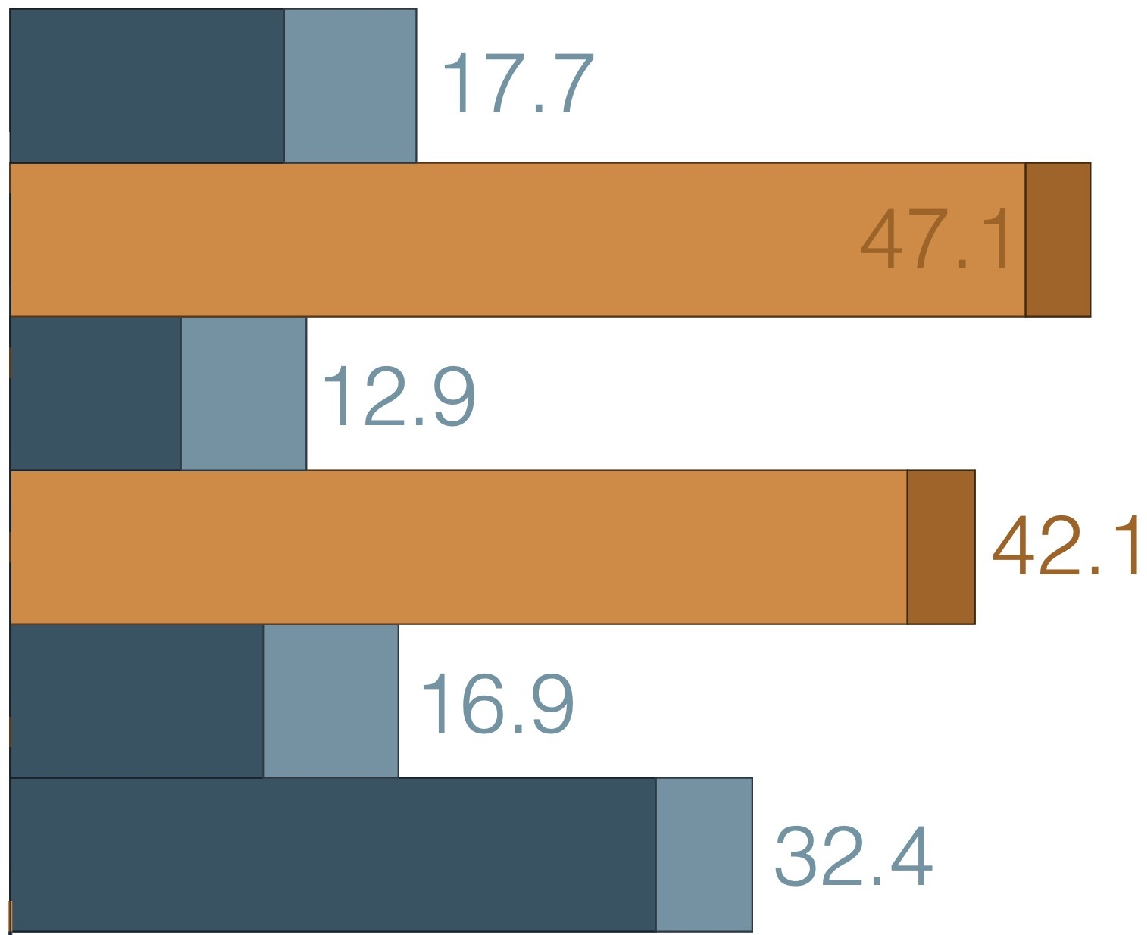}} \\
\multicolumn{1}{c}{MO-Specific Parameters} & & & & & & & \\ \cmidrule(lr){1-1}
DFT + U$^d$ & -2.60 & -43.53 & -1.84 & -30.79 & -1.72 & -29.27 & \\
DFT + U$^{d,p}$ & -18.01 & -41.77 & -12.67 & -29.38 & -11.96 & -28.28 & \\
DFT + U$^d_\textrm{eff}$ & -10.19 & -45.89 & -7.35 & -33.09 & -6.54 & -31.29 & \\
DFT + U$^{d,p}_\textrm{eff}$ & -40.63 & -46.36 & -29.15 & -33.26 & -26.75 & -31.79 & \\
DFT + U$^d$ + J$^d_{\textrm{z}}$ & -19.24 & -44.87 & -13.73 & -32.01 & -12.39 & -30.40 & \\
DFT + U$^{d,p}$ + J$^{d,p}_{\textrm{z}}$ & -49.96 & -45.16 & -35.48 & -32.07 & -32.83 & -30.81 & \multirow{-7}{*}{\includegraphics[width=0.21\textwidth]{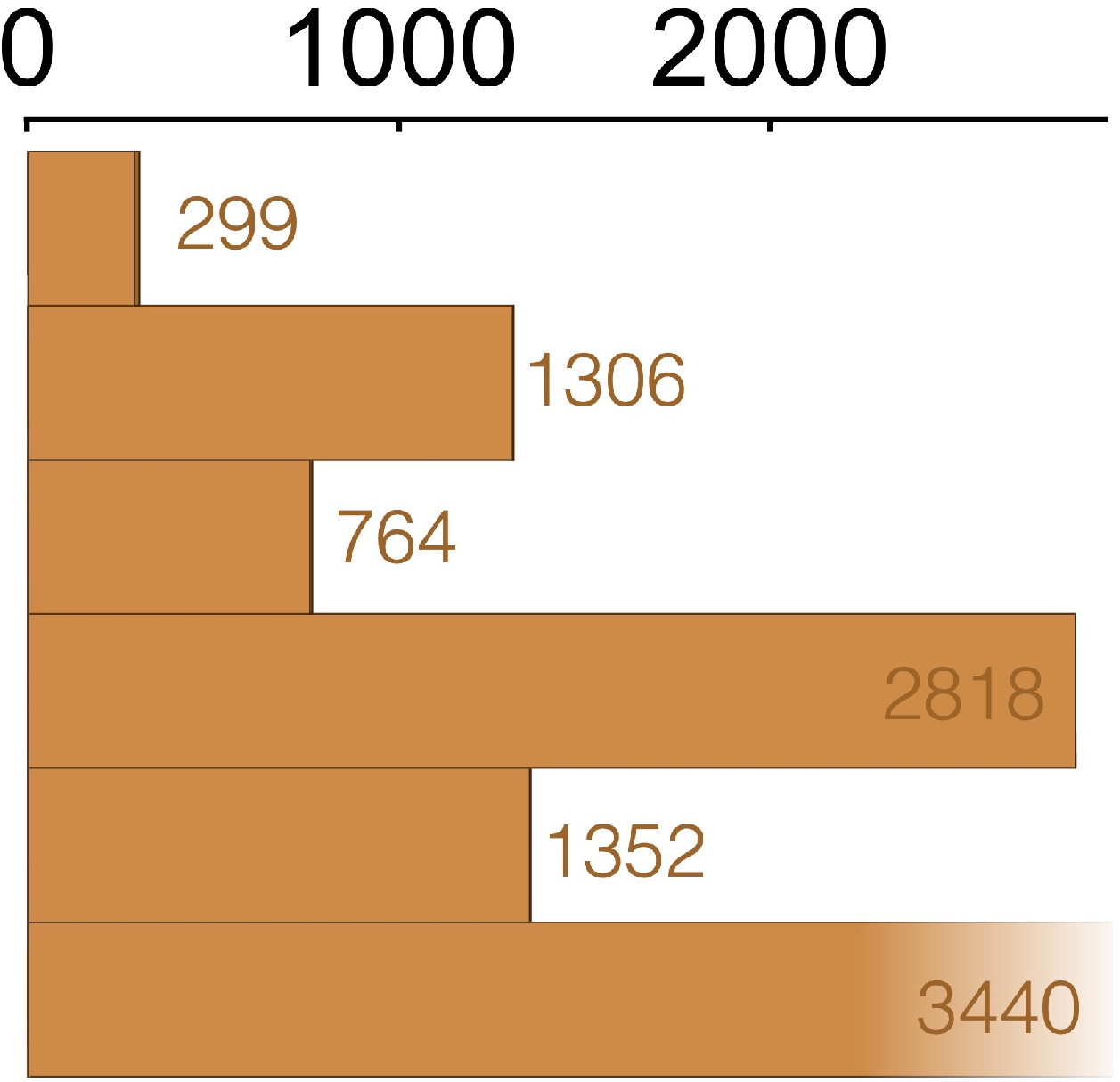}} \\
\multicolumn{1}{c}{Experiment} & & & & & & & \\ \cmidrule(lr){1-1} \cmidrule(lr){2-3}
{INS $\vert$ Hutchings (1972)}\cite{Hutchings_1972} & 0.69 & -9.50 & \multicolumn{4}{c}{} & \\
{Thermo $\vert$ Shanker (1973)}\cite{Shanker1973} & -0.69\footnote{\edit{Ref. \citen{Shanker1973} reports a preference for antiferromagnetic alignment for both $J_1$ and $J_2$.}} & -8.66 & \multicolumn{4}{c}{} & \\
{Magnon $\vert$ Lockwood (1992)}\cite{Lockwood1992} & 0.69 & -9.61 & \multicolumn{4}{c}{} & \\
{Magnon $\vert$ Dietz  (1971)}\cite{Dietz1971} & | & -9.18 & \multicolumn{4}{c}{} &  \\
\end{tabular}
\end{center}
\caption{Comparison of exchange coupling parameters from various Hubbard functionals, where \emph{U$_\textrm{eff}=\textrm{U}-\textrm{J}_\textrm{z}$}. The ``\emph{AF$_\textrm{II}$ NiO Parameters}" header refers to functionals tested using the Hubbard parameters arising from linear response on the ground state \emph{AF$_\textrm{II}$} magnetic ordering of \emph{NiO} (i.e., column 4 in Table \ref{Final_HPs+Errors}). ``\emph{MO-Specific Parameters}" refers to the use of the Hubbard parameters pertaining specifically to each magnetic ordering (i.e., \emph{FM} HPs applied to \emph{FM} ordering via Hubbard functional, \emph{AF$_\textrm{I}$} parameters applied to \emph{AF$_\textrm{I}$} ordering, etc.). Histogram shows  minimum (leftmost bar), range (stacked bar) and maximum (digit) of all \emph{\% RMS errors} with respect to $J_1$, $J_2$ pairs from each of the four experiments listed above. Color of bar corresponds to  spin parametrization method used to obtain that best $J_1$, $J_2$ pair (i.e., tan for \emph{Method A}, blue for \emph{Method B} and orange for \emph{Method C}).}
\label{FunctionalsVsExchange}
\end{table*}
%%%%%%%%%%%%%%%%%%%%%%%%%%%%%%%%%%%%%%%%%%%%%%%%%%%%%%

\subsection{\label{Results_interatomic_spin}\edit{Comparison of spin parametrization methods for the Heisenberg model mapping}}

In Table \ref{FunctionalsVsExchange}, we resolve our $ \vert \textbf{S}^\textrm{AS} \vert$ results to the level of the individual Hubbard functional and  spin parametrization method. Principally, we report obtained values for $J_1$ and $J_2$. Secondarily, for visual comparison, the final column  shows the range of RMS errors obtained by comparing the best pair of coefficients from that functional to each of the four experiments listed at the bottom of the table. To clarify, for each functional, we take the best $J_1$,$J_2$ pair of the three spin parametrization methods and calculate their percent RMS error with respect to each of the four experiments via Eq. (\ref{percentRMSEJ1J2}). We then report the minimum (leftmost bar), range (stacked bar) and maximum (digit) of all those errors to compare across other functionals. The color of the bar corresponds to the  spin parametrization method used to obtain that best ($J_1$,$J_2$) pair (i.e., tan for Method A, blue for Method B and orange for Method C).

We prepare an identical error analysis scheme in Table \ref{Past_Works_Comparison}, where we compare the exchange interaction parameters $J_1$ and $J_2$ reported in the literature to our own results, obtained from the various  spin parametrization methods proposed in Section \ref{HeisenbergModel}. For values reported in the literature, we reverse engineered the exchange interaction parameters for Methods B and C based on those reported for Method A (the conventional method) in addition to the reported values for the Ni magnetic moment for each configuration. 

\edit{It should be clarified that we use RMS error in Tables \ref{SpinMomentErrors}, \ref{FunctionalsVsExchange}, and \ref{Past_Works_Comparison} as a measure of proximity to the reported values of the four different experiments cited. The RMS error is not asserting the proximity of our computed parameters to the true, physical values of these parameters (e.g., a \% RMS error of $20\%$ on $J_1=0.7$ meV does not mean we claim to know the value of $J_1$ to within $\pm 0.14$ meV on the basis of calculations.}

All Hubbard functionals we tested produced exchange coupling parameters that are highly competitive with those from other investigations. Generally, however, the literature reports a wide range of values for this particular magnetic property, in part due to large variation in calculation of the magnetic moments (approximately half use some form of the total spin density difference integrals, others use atom-centered sphere integration up to inconsistent maximum radii; all use $ \textrm{S}_z$). Only a few computations use PAW, and many calculate the exchange coefficients with methods other than that involving total energy differences. It is crucial to recognize that the variation in spin moment integrations severely limits the reliability of any comparison across works of literature, and it is these incongruities in procedure that reinforce our inclination to benchmark with experiment. 

Now for the analysis; referencing Tables \ref{SpinMomentErrors} and \ref{FunctionalsVsExchange}, for converged total energies and their corresponding magnetic moments, we find that spin parametrization Methods B and C generally predict exchange interaction values smaller in magnitude, and closer to experiment, than the method that uses idealized values for Ni magnetization. This much is also demonstrated by the curves (the exchange parameters as functions of AF$_\textrm{II}$ U$^d$ variation) in Figures \ref{NiUvsMagJs} (c) and (d) as U$^d$ approaches its in situ values of U$^d=5.58$ eV and U$^d_\textrm{eff}=5.11$ eV.

%%%%%%%%%%%%%%%% Past Works Comparison %%%%%%%%%%%%%%%%%%
\begin{table*}
\begin{center}
\centering
\begin{tabular}{r|cc|cc|cc|cc|c}
  \multicolumn{1}{c}{{\color[HTML]{FFFFFF} }} &  & \multicolumn{1}{c}{{\color[HTML]{FFFFFF} }} & \multicolumn{2}{c}{\color{tan} Method A} & \multicolumn{2}{c}{\color{navy} Method B} & \multicolumn{2}{c}{\color{copper} Method C} & \\ \cmidrule(lr){4-5}\cmidrule(lr){6-7}\cmidrule(lr){8-9}
\rowcolor[HTML]{FFFFFF}[\overhang]
\multicolumn{1}{c}{} &  &  & \multicolumn{2}{c|}{{\tiny\tcw\cellcolor{tan} \spc $\textrm{S}_\textrm{FM}=\textrm{S}_{\textrm{AF}_{\textrm{I}}}=\textrm{S}_{\textrm{AF}_{\textrm{II}}}=1$} \spc} & \multicolumn{2}{c|}{{\tiny\tcw\cellcolor{navy}\spc $\textrm{S}_\textrm{FM}=\textrm{S}_{\textrm{AF}_{\textrm{I}}}=\mathbf{\textrm{S}_{\textrm{AF}_{\textrm{II}}}}$} \spc} & \multicolumn{2}{c|}{{\tiny\tcw\cellcolor{copper}\spc $\textrm{S}_\textrm{FM}\neq\textrm{S}_{\textrm{AF}_{\textrm{I}}}\neq\textrm{S}_{\textrm{AF}_{\textrm{II}}}$} \spc} & \\ \rowcolor[HTML]{FFFFFF}[\overhang]
& U & J$_{\textrm{z}}$ & {\tcw\cellcolor{tan} $J_1$} & {\tcw\cellcolor{tan} $J_2$} & {\tcw\cellcolor{navy} $J_1$} & {\tcw\cellcolor{navy} $J_2$} & {\tcw\cellcolor{copper} $J_1$} & {\tcw\cellcolor{copper} $J_2$} & \% RMS Error \\
\rowcolor[HTML]{FFFFFF}[\overhang]
\multirow{-1}{*}{$^\textrm{Ref}\vert$ XC} & \spc(eV) & (eV)\spc & {\tcw\cellcolor{tan}\spc\spc\spc (meV)\spc\spc\spc\spc} & {\tcw\cellcolor{tan}(meV)} & {\tcw\cellcolor{navy}\spc (meV) \spc\spc} & {\tcw\cellcolor{navy}(meV)} & {\tcw\cellcolor{copper}\spc\spc (meV) \spc\spc\spc} & {\tcw\cellcolor{copper}(meV)} & \includegraphics[width=0.1875\textwidth]{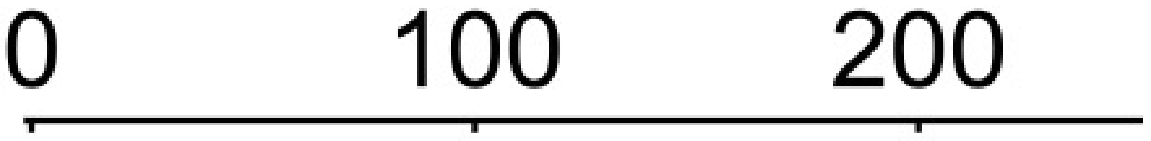} \\ \hline
PBE &  &  & 8.23 & -29.32 & 9.73 & -34.68 & 1.37 & -30.48 & \\
PBE+U$^d$ & 5.58 & & 0.95 & -10.20 & 0.67 & -7.22 & 0.60 & -6.95 & \\
PBE+U$^d_\textrm{eff}$ & 5.58 & 0.47 & 1.03 & -11.10 & 0.74 & -8.00 & 0.66 & -7.68 & \\
PBE+U$^d$+J$^d_{\textrm{z}}$ & 5.58 & 0.47 & 0.96 & -10.24 & 0.69 & -7.31 & 0.62 & -7.04 & \\ \cline{1-9}
\cite{logemann2017}$\vert$ PBE+U & 6.30 & 1.00 & \textbf{1.10} & \textbf{-11.90} & 1.66 & -17.96 & 2.05 & -17.83 & \\
\cite{archer_exchange_2011}$\vert$ PBE &  &  & \textbf{0.60} & \textbf{-22.25} & 1.32 & -48.95 & 2.32 & -44.02 & \\
\cite{archer_exchange_2011}$\vert$ HSE &  &  & \textbf{1.15} & \textbf{-10.5} & 1.69 & -15.46 & 1.34 & -14.52 & \\
\cite{archer_exchange_2011}$\vert$ PSIC &  &  & \textbf{1.65} & \textbf{-12.35} & 2.55 & -19.10 & 2.37 & -18.40 & \\
\cite{kodderitzsch2002}$\vert$ SIC-LSD$^*$ &  &  & \textbf{1.15} & \textbf{-6.0} & 1.89 & -9.88 & 1.53 & -9.12 & \\
\cite{zhang_pressure_2006}$\vert$ GGA+U & 6.30 & 1.00 & \textbf{0.87} & \textbf{-9.54} & 1.23 & -13.55 &  & &  \\
\cite{moreira2002}$\vert$ LDA &  &  & \textbf{5.95} & \textbf{-35.65} & 17.13 & -102.65 &  & &  \\
\cite{moreira2002}$\vert$ UHF &  &  & \textbf{0.4} & \textbf{-2.3} & 0.44 & -2.53 &  & &  \\
\cite{moreira2002}$\vert$ Fock-35 &  &  & \textbf{0.95} & \textbf{-9.85} & 1.24 & -12.90 &  & &  \\
\cite{moreira2002}$\vert$ Fock-50 &  &  & \textbf{0.75} & \textbf{-5.3} & 0.92 & -6.49 &  & &  \\
\cite{moreira2002}$\vert$ B3LYP &  &  & \textbf{1.2} & \textbf{-13.35} & 1.73 & -19.19 &  & &  \\
\cite{fischer2009}$\vert$ LSIC &  &  & \textbf{1.42} & \textbf{-6.95} & 2.02 & -9.87 &  & &  \\
\cite{fischer2009}$\vert$ LSIC$^*$ &  &  & \textbf{0.15} & \textbf{-6.92} & 0.21 & -9.83 &  & &  \\
\cite{kvashnin2015}$\vert$ LSDA$^*$ &  &  & \textbf{0.54} & \textbf{-21.50} & 0.64 & -25.18 &  & &  \\
\cite{kvashnin2015}$\vert$ LSDA+U$^*$ & 8.2 & 0.95 & \textbf{-0.03} & \textbf{-6.80} & -0.03 & -7.97 &  & &  \\
\cite{kvashnin2015}$\vert$ LSDA+DMFT$^*$ &  &  & \textbf{-0.04} & \textbf{-6.53} & -0.05 & -7.65 &  & & \\
\cite{kotani2008}$\vert$ LDA$^*$ &  &  & \textbf{0.3} & \textbf{-28.3} & 0.82 & -77.50 &  &  & \\
\cite{Gopal2017}$\vert$ PBE+U$^{d,p}$ & 7.63\footnote{Applied U$^d=$ 7.63 eV and U$^p=$ 3.00 eV.} & & \textbf{2.36} & \textbf{-25.15} & 4.25 & -45.42 &  &  & \\
\cite{Gopal2017}$\vert$ LDA+U$^{d,p}$ & 7.49\footnote{Applied U$^d=$ 7.49 eV and U$^p=$ 2.60 eV.} &  & \textbf{1.48} & \textbf{-20.04} & 4.05 & -54.86 &  &  & \\
\cite{jacobsson2013}$\vert$ LDA+U & 8.0 & 0.95 & \textbf{2.4} & \textbf{-28.0} &  &  &  &  & \\
\cite{korotin2015}$\vert$ LDA+U$^*$ & 8.0 & 0.9 & \textbf{0.2} & \textbf{-9.4} &  &  &  &  & \multirow{-25}{*}{\includegraphics[width=0.1875\textwidth]{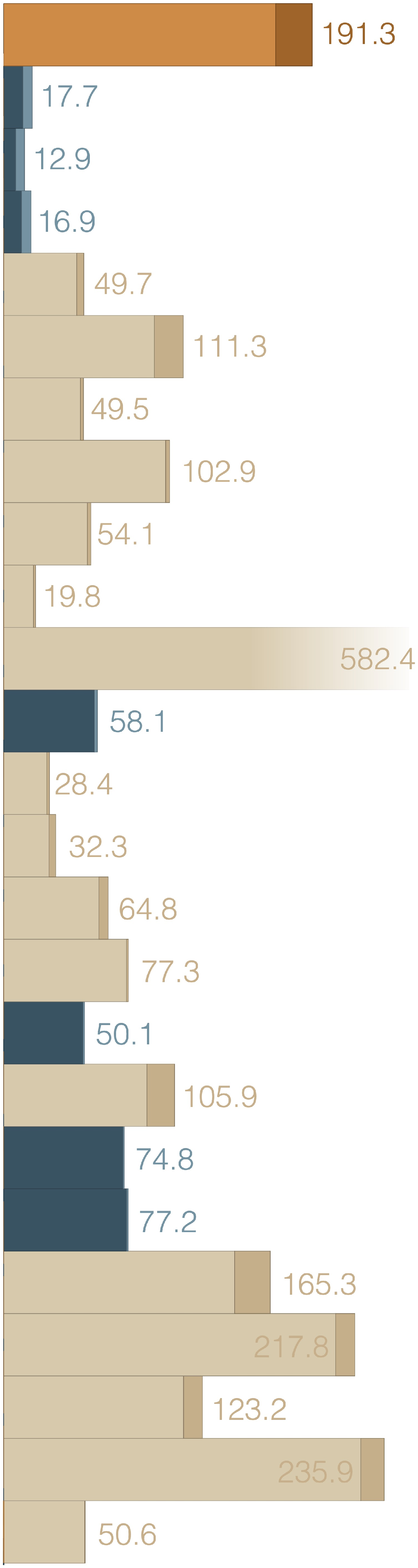}} \\
\multicolumn{1}{c}{{\color[HTML]{FFFFFF} }} &  & \multicolumn{1}{c}{{\color[HTML]{FFFFFF} }} & \multicolumn{2}{c}{\color{sea} Experiment} & \multicolumn{2}{c}{} & \multicolumn{2}{c}{} & \\ \cmidrule(lr){4-5}

\multicolumn{3}{r|}{\cite{Hutchings_1972}$\vert$ INS} & \textbf{0.69} & \multicolumn{1}{c|}{\textbf{-9.50}} & \multicolumn{5}{c}{} \\
\multicolumn{3}{r|}{\cite{Shanker1973}$\vert$ Thermo} & \textbf{-0.69} & \multicolumn{1}{c|}{\textbf{-8.66}} & \multicolumn{5}{c}{}  \\
\multicolumn{3}{r|}{\cite{Lockwood1992}$\vert$ Magnon} & \textbf{0.69} & \multicolumn{1}{c|}{\textbf{-9.61}} & \multicolumn{5}{c}{}  \\
\multicolumn{3}{r|}{\cite{Dietz1971}$\vert$ Magnon} & | & \multicolumn{1}{c|}{\textbf{-9.18}} & \multicolumn{5}{c}{} \\
\end{tabular}
\end{center}
\caption{Comparison of exchange interaction parameters from different XC functionals (or experiment) across literature works for each spin-parametrization method. Parameters for \emph{Methods B} and \emph{C} are reverse engineered from past works based on reported values for MO-dependent magnetization on \emph{Ni} atoms and $J_1$ and $J_2$ values, translated according to Hamiltonian conventions outlined in Table \ref{Heisenberg_Conventions}. Reported values from each work are in bold. Histogram shows minimum (leftmost bar), range (stacked bar) and maximum (digit) of all \emph{\% RMS errors} with respect to $J_1$, $J_2$ pairs from each of the four experiments listed. Color of bar corresponds to  spin parametrization method used to obtain that best $J_1$, $J_2$ pair (i.e., tan for \emph{Method A}, blue for \emph{Method B} and orange for \emph{Method C}). Asterisk (*) indicates that these exchange parameters were calculated via a method other than that using total energy (i.e., Green's Functions).}
\label{Past_Works_Comparison}
\end{table*}
%%%%%%%%%%%%%%%%%%%%%%%%%%%%%%%%%%%%%%%%%%%%%%%%%%%%%%%%%

Table \ref{FunctionalsVsExchange} shows that Method C routinely predicts exchange parameters smaller in magnitude than those from Methods B and A. This may be attributed to the presence of and a heavy reliance on a strong magnetic moment on FM NiO---that, or the AF$_\textrm{II}$ moment is comparatively too weak. Whatever the explanation, the message is clear; if the spin moments are large enough, then it is well worth the effort to weight the total energies with them, as spin parametrization Methods B and C predict exchange parameters more closely in line with experiment. That much is also evident in the \% RMSE in Row 1 of Table \ref{SpinMomentErrors}. The same AS data demonstrate a slight preference for Method B over Method C, however, the difference is minute.

The same conclusions cannot be drawn from Table \ref{Past_Works_Comparison}. While $J_2$ is consistently smallest in magnitude under Method C, $J_1$ actually performs best under Method A. This is consistent with our own data, recalling that the literature  analyzed here exclusively calculates the magnetic moment as a projection on the z-axis, whereas values from our own Hubbard functionals reported in Table \ref{FunctionalsVsExchange} use $ \textbf{S}^\textrm{AS} $ as per the conclusions of Section \ref{Results_intra_spin_moment}. If we better align our calculations with the literature by using $ \textrm{S}_z^\textrm{AS} $, we can see from Table \ref{SpinMomentErrors} that indeed, Method A minimizes the error compared to the other spin parametrization methods. This demonstrates the sensitivity of the interplay between the magnetic moments and total energies of each MO of NiO. Under our preferred spin treatment procedure, our results indicate that Methods B and C are, somewhat interchangeably, the most reliable in obtaining realistic exchange interaction parameters for NiO.

\subsection{\label{Hubbard Functionals}\edit{Effect of MO-specific 
Hubbard parameter calculation}}

Going back to Table \ref{FunctionalsVsExchange}, we see the uncorrected PBE XC functional predicts, at best, $J_1=1.37$ meV under Method C and $J_2=-29.32$ meV under Method A, values which are far from corroborated by our experimental benchmarks of $J_1=0.69$ meV and $J_2=-9.50$ meV.\cite{Hutchings_1972} The error arising from parametrization schemes B and C can be attributed in part to PBE's incorrect prediction of the Ni magnetic moments under the AS population analysis ($\mu_\textrm{FM}= 1.09\ \mu_B$, $\mu_{\textrm{AF}_\textrm{I}}= 1.56\ \mu_B$, $\mu_{\textrm{AF}_\textrm{II}}= 1.30\ \mu_B$).

\edit{PBE nonetheless correctly predicts} the preference for ferromagnetic alignment in $J_1$ and antiferromagnetic alignment in $J_2$, unlike all functionals incorporating MO-specific parameters (center of Table \ref{FunctionalsVsExchange}). These performed poorly, not only quantitatively---note the inflated \% RMS error scale, dilated by two orders of magnitude---but qualitatively, as they failed to predict the correct magnetic alignment for $J_1$. This decisively answers the query posed earlier; while the U and \editt{J$_\textrm{z}$} for a particular subspace in FM NiO differ from those calculated for other MOs (see Tables \ref{dmatpuopt_options} and \ref{Final_HPs+Errors}), the total energies resulting from the application of these magnetic ordering-specific parameters to their corresponding system are not necessarily comparable to one another. We must emphasize, however, that this will not necessarily be the case for any and all Hubbard functionals; there remain quite a few functionals untested in this article, and further testing is needed to determine if this ineptitude is an intrisinc property of Hubbard corrective protocol.

This conclusion is somewhat unexpected. The MO of a particular material system influences the magnitude of its atomic magnetic moments, as is demonstrated in Figures \ref{NiUvsMagJs} (a) and (b). These moments are direct functions of valence subspace occupations, and so changing the magnetic moments in turn alters the amount of self-interaction intrinsic to each subspace and, therefore, the Hubbard parameters and the efficacy of their host Hubbard functionals. If the total energies of each system are individually more accurate by application of a Hubbard corrective functional, then is it not expected that the total energy differences between these systems become more accurate, as well?

The linearity condition of linear response is based heavily on the quadratic nature of SIE (i.e., the deviation from piece-wise linearity of the total energy with respect to fractional charge). It is worth noting, however, that the  response behavior in the first-principles calculation of the Hubbard parameters for certain Ni \textit{3d} subspaces---notably upon application of the beta perturbation to the FM and AF$_\textrm{I}$ MOs---noticeably deviated from linearity. This non-linear response suggests that the SIE in these systems may require a correction of order higher than quadratic, possibly as a result of the added complexity in enforcing the system's non-ground state MOs.

\edit{We might be tempted to conclude then} that applying different sets of Hubbard parameters to these systems is akin to changing its functional entirely, to the degree that the resulting total energies are no longer comparable (i.e., asserting that a DFT+U$^d=5.58$ eV functional is inherently different to a DFT+U$^d=6.91$ eV functional). However, this justification is not compatible with our understanding of the Hubbard parameters as intrinsic, ground state properties of the degree to which SIE and SCE infect the system. In maintaining the integrity of this understanding, then, it is more plausible that we have simply not found/tested a Hubbard-like functional adequately equipped to deal with long-range magnetic ordering of this nature. In applying the same Hubbard parameters inside this particular set of functionals to all structures, its possible that the results benefit from a black-box cancellation of errors that manifests in the total energy difference domain. Whatever the case, it is clear that calculating and applying MO-ordering specific Hubbard parameters is a disadvantageous complexity for deriving the exchange coupling parameters for magnetic NiO, at least for the six Hubbard functionals tested here.

\subsection{\label{Hubbard Functionals2}\edit{Effect of Hubbard functional choice}}

Whatever ails the MO-specific functionals does not affect the Hubbard functionals employing only the AF$_{\textrm{II}}$ NiO parameters, in which the same Hubbard functionals are applied to all structures regardless of MO. Like PBE, these functionals correctly predict the preference for ferromagnetic alignment in $J_1$ and antiferromagnetic alignment in $J_2$. This remains true as long as the Hubbard parameters are reasonable in magnitude, as demonstrated in Figures \ref{NiUvsMagJs} (c) and (d); under the standard DFT+U$^d$ functional, both the NN and the NNN exchange interactions gradually weaken with increasing U$^d$, coinciding with experimental values around the first-principles value of U$^d$.

In particular, the DFT+U$^d_\textrm{eff}$ yields $J_1$ and $J_2$ coefficients strikingly in line with experimental values under the Method B spin parametrization, contributing to shared RMS errors ranging from 7.4\% to 12.9\%. It is closely followed in performance by the same functional under Method C spin parametrization (8.6-14.5\% error), and then by the Liechtenstein DFT+U$^{d}$+J$^{d}_\textrm{z}$ functional---in which we apply both HPs to the Ni \textit{3d} subspace only--under Method B (11.0-16.9\% error). This ranking remains stable even as the intra-atomic spin population analysis technique varies, although only when considering $\vert \textbf{S} \vert$. \edit{When using instead} $\textrm{S}_z$, the Liechtenstein DFT+U$^{d}$+J$^{d}_\textrm{z}$ under Method A performs best.

The $J_1$, $J_2$ pairs from these Hubbard functionals are highly competitive with those from other semi-local or hybrid functionals reported in the literature. Looking back at Table \ref{Past_Works_Comparison}, the best DFT prediction for $J_1$ on NiO remains that from Fock-35 results of Ref.~\citen{moreira2002}, with an individual parameter RMS error of 8.7\%. For $J_2$, the best predictions are held by the GGA+U$_\textrm{eff}=5.3$ eV calculation of Ref. \citen{zhang_pressure_2006} ($J_2$ RMS errors ranging from 0.4\% to 10.2\%) and the LDA+U$^d_\textrm{eff}=7.1$ eV conducted in Ref. \citen{korotin2015} ($J_2$ errors 1.1-8.6\%). In terms of pair-wise performance, however, the in situ Hubbard functionals monopolize the best error minimization rankings, reflecting positively on the error compensation and spin parametrization techniques. At minimum, this demonstrates that the Hubbard functionals are competitive with the more costly hybrid functionals in achieving total energy difference results that reflect experiment.

\edit{\subsection{\label{Hubbard Functionals Explain}\edit{Mechanisms of Hubbard functional choice}}}

\edit{It is possible to rationalize the dependence of $J_1$ and $J_2$ on the choice of Hubbard functional and parameter strength in a relatively simple way. To see this, we start with the energy expressions for Method B, Eqs. (\ref{J1wmagnsimple}) and (\ref{J2wmagnsimple}). Taking the former, $J_1$, we first note that a sufficient U$^{d}$ will change the FM and AF$_\textrm{I}$ systems from metallic to insulating. This will tend to reduce the relevant total energy difference; the total energy of the metallic state depends significantly, through the kinetic energy, on the Fermi densities of the two systems, whereas in the insulating systems, it will primarily depend on weaker differences of exchange and correlation effects. As we can see from Figure \ref{NiUvsMagJs} (a), the local moments become more similar in the two states at larger U values, which moderates the tendency for the different Hubbard energy contributions to increase. As these local moments increase with U$^{d}$, furthermore, so too the denominator in Eq. (\ref{J1wmagnsimple}) increases quadratically, which further brings down the $J_1$. Thus, we can visualize how increasing U$^{d}$ tends to bring down the predicted $J_1$. Of course, that effect is moderated in the Dudarev scheme when U$^{d}$ is reduced to U$^d_\textrm{eff}$ by an increasing J$^d_\textrm{z}$.}

\edit{Meanwhile, it is interesting that the exchange parameters arising from DFT+U$^d$ and the Liechtenstein DFT+U$^{d}$+J$^{d}_\textrm{z}$ are so similar. In fact, the similarity between the exchange parameters arising from either functional is tenacious, remaining even as we vary U$^{d}$ and J$^{d}_\textrm{z}$ (keeping the first-principles ratio between U$^{d}$ and J$^{d}_\textrm{z}$ constant). The relative agnosticism of the \editt{inter-atomic, Heisenberg} exchange parameters to the addition of \editt{a non-zero J$_\textrm{z}$ parameter} suggests that the spin localization facilitated by the U$^{d}$ term is already quite robust. The addition of U$^p_\textrm{eff}$ largely mitigates the effect of U$^d_\textrm{eff}$ on $J_1$, and not in a beneficial way when using the first-principles parameters. We return to discuss this effect further in the next section.}

\edit{As for the substantial moderation of $J_2$ by the addition of U$^{d}$, the same arguments and trends largely hold. The total energy differences in Eq. (\ref{J2wmagnsimple}) are reduced when all three systems are insulating. Once again, as the moments are increased with increasing U$^d_\textrm{eff}$, so the $J_2$ decreases due to the quadratic expression in the denominator. This effect is weakened by J$^d_\textrm{z}$ in the Dudarev functional, by definition, and beneficially so when $J_2$ is compared against experiment. The effects of J$^d_\textrm{z}$ and J$^p_\textrm{z}$ are much more weak in the Liechtenstein functional, where it is  also more difficult to interpret. The main difference in the trends observed between $J_1$ and $J_2$ is a different sign in the change with the addition of finite U$^p_\textrm{eff}$, so that increasing U$^p_\textrm{eff}$ reduces the magnitude of $J_2$, as we go on to discuss further. Once again, however, the addition of the first-principles U$^p_\textrm{eff}$ is detrimental to the quality of $J_2$.}\\

%%%%%%%%%%%%%%%%%%%% Up Variation DOS %%%%%%%%%%%%%%%%%%%%%%%%%%%%
\begin{figure*}
    \centering
    \includegraphics[width=1.0\textwidth]{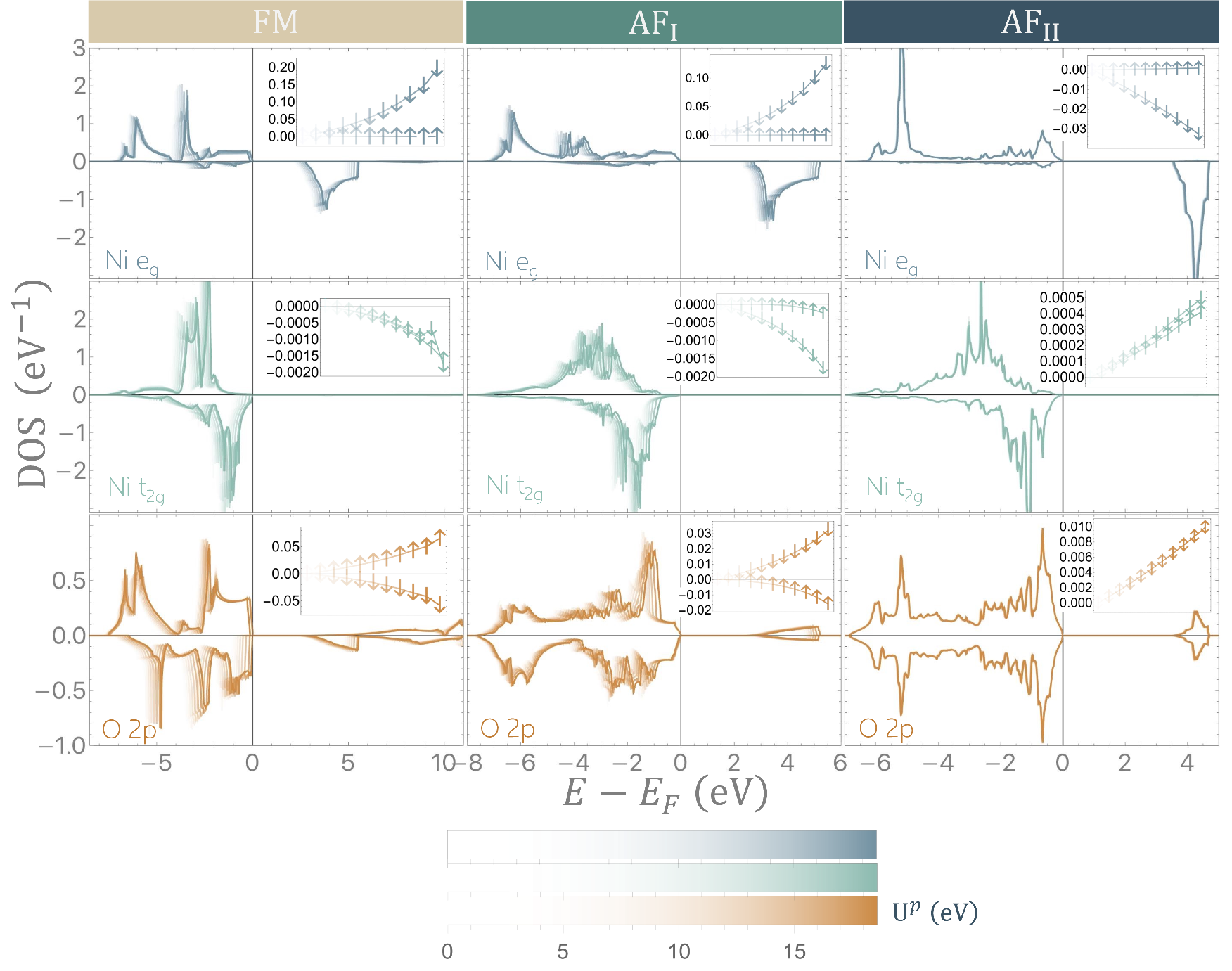}
    \caption{\edit{Overlaid atomic (AE) orbital projected density of states (pDOS) on two particular atoms of the four-atom unit cell, resolved by crystal field splitting characterization on \emph{Ni} (e$_g$ blue, top row; t$_{2g}$ light green, middle row) and \emph{O} (\textit{2p} orbital orange, bottom row) for each magnetic ordering (\emph{FM} left column; \emph{AF$_\textrm{I}$} middle column; \emph{AF$_\textrm{II}$} right column) of \emph{NiO}. Positive values indicate majority spin DOS, negative values indicate minority spin. Only the component within the PAW augmentation sphere cutoff $r_c$ for each species is included, with no interstitial contributions, hence the relatively sharp (atomic-like) features of the pDOS. Moreover, only the conduction band edge position should be interpreted here, as the Kohn-Sham basis is not thought to be complete for the conduction band for all MOs in the energy range shown. Insets show corresponding relative changes in majority spin \emph{($\sigma=\uparrow$)} and minority spin \emph{($\sigma=\downarrow$)} occupancies $\Delta n^\sigma=100(n^\sigma \vert_{\emph{U}^p} - n^\sigma \vert_{\emph{U}^p=0.45})/n^\sigma \vert_{\emph{U}^p=0.45}$ as a function of \emph{U$^p$ [eV]} for that particular MO and crystal field characterization. Opacity of plot is a function of \emph{U$^p$ (eV)} in the Dudarev et al. \emph{DFT+U$^{d,p}$} functional, where \emph{U$^d=5.58$ eV} is kept constant and applied to all MOs. Variations in the \emph{AF$_\textrm{II}$} case are present but barely discernible.}}
    \label{UpVariation}
\end{figure*}
%%%%%%%%%%%%%%%%%%%%%%%%%%%%%%%%%%%%%%%%%%%%%%%%%%%%%%%%%%

\subsection{\label{Oxygen}Effect of Hubbard Parameters on Oxygen}

Introduction of HPs on the oxygen \textit{2p} orbitals, across all Hubbard functionals, seems to have the effect of strengthening the NN ferromagnetic exchange while simultaneously weakening the antiferromagnetic coupling between layers, similar to how HSE and PSIC perform compared to raw PBE.\cite{archer_exchange_2011} For example, relative to DFT+U$^{d}$+J$^{d}_\textrm{z}$, the additional U$^{p}$ and J$^p_\textrm{z}$ in DFT+U$^{d,p}$+J$^{d,p}_\textrm{z}$ increase $J_1$ to 1 meV, but bring $J_2$ down in magnitude to $-7.35$ meV. These are typically unwelcome modifications under Methods B and C, for both $J_1$ and $J_2$, since the latter is chronically underestimated by a large margin and the former hovers tightly around the experimental value of $0.69$ eV.

Nonetheless, the regularity of the phenomenon is intriguing and worth some discussion. In the much simplified picture of Eq. (\ref{totalenergyheis}), the spin dependence of the AF$_\textrm{II}$ total energy is informed exclusively by $J_2$ interactions, and the FM total energy is informed heavily, although not exclusively, by $J_1$ interactions. Moreover, as discussed in Section \ref{ExchangeParameters}, the crystal field splitting of orbitals is heavily implicated in the direct and superexchange mechanisms that collectively give rise to interatomic exchange. We can perhaps look for clues in the FM and AF$_\textrm{II}$ NiO density of states (DOS) plots, shown in Figure \ref{UpVariation} as functions of U$^p$, keeping U$^d=5.58$ eV constant.

In FM NiO, the occupied minority spin states on oxygen and the unoccupied minority spin states on Ni are shifted higher in energy with increasing U$^{p}$, although the former at a faster rate than the latter. By itself, this means we should expect the exchange energy, and thus $J_1$, to gradually reduce in magnitude. However, simultaneous to this feature, we notice two pivotal details about the shifting occupations; (1) the $t_{2g}$ states are decreasing in occupation, so the Coulombic repulsion that contributes to a strong AF exchange interaction is weakening; and (2) the oxygen $2p$ orbitals are increasing in occupation, as demonstrated by the gradually inflating area underneath the pDOS curve as well as the \edit{inset graphs in Figure \ref{UpVariation}}. Thus, the intra-atomic Coulomb repulsion grows, thereby strengthening the FM preference of the superexchange mechanism. Because intra-atomic exchange overwhelms its interatomic counterpart by several orders of magnitude, we expect these local oxygen dynamics to eclipse the gradual reduction in exchange energy coming from the evolution of the minority spin pDOS. Collectively, these justifications point to an overall increasing FM preference for $J_1$.

For the AF$_\textrm{II}$ NiO case, the Ni and O subspace occupancies both become more integer-like upon application of HPs on O in addition to those already applied on Ni---more integer-like and, thus, more localized spatially. As the orbitals cling more closely to their respective nuclei, there is less overlap between neighboring atomic orbitals. In the NNN case, less orbital overlap translates to a decrease in virtual hopping and therefore a weaker exchange bridge; by consequence, $J_2$ decreases in magnitude. Yet, unlike in 180$^\circ$ superexchange, the 90$^\circ$ superexchange is mediated by the Coulombic exchange between orthogonal orbitals on the same oxygen atom (see Figure \ref{Superexchange_Schematic}). As orthogonal orbitals on the same atom become more localized, the exchange increases in magnitude, thereby intensifying the ferromagnetic preference embodied by $J_1$. So, $J_1$ increases in magnitude and $J_2$ decreases in magnitude, precisely as observed. In this way, it is possible that the application of the oxygen HPs on top of those applied on Ni serves to partially relocalize the atomic orbitals.

In Figure \ref{UpVariation}, it is interesting that the AF$_\textrm{II}$ pDOS character varies almost imperceptibly with changes in oxygen's U$^p$, which changes only $E_F$ in a manner rigidly linear with the magnitude of the Hubbard parameter. We mention this to preemptively assuage concerns that it belies a plotting error and tentatively propose that the phenomenon is related to the fact that the oxygen atoms in AF$_\textrm{II}$ NiO have absolutely no net magnetic moment. \edit{When we set U$^p$ to a large, finite number, the DOS undergoes significant changes.}

\edit{We emphasize here that the pDOS shown in Figure 5, in addition to SM Figures 2 and 3, are truncated within the PAW augmentation spheres, with no interstitial contribution, based on the PAW population analysis. It therefore exhibits sharper, more atomic-like features than would be observed experimentally. An increased smearing would be needed for comparison to experiment. Furthermore, only the conduction band edge should be interpreted, and features no higher in energy thereof, due to insufficient basis resolution. Our principle objective here is to understand the mechanisms of inter-atomic exchange interactions.}

\edit{Despite these limitations, our AF$_\textrm{II}$ NiO pDOS does not deviate significantly in its features from those of prior DFT+U calculations.\cite{Trimarchi2018,Han2006,Naz2018,Egbo2020,Dawson2015,zhang2008} The upper band is clearly of primarily Ni $e_g$ character, whereas the lower band edge is characterized by the O $2p$, followed by non-negligible contributions of Ni $t_{2g}$ and $e_g$ character. The failure of DFT+U-like methods to replicate other features of the experimental pDOS is well-documented across the literature. For example, our oxygen $2p$ spectrum exhibits a two-peaked structure (\editt{with edges at} -5.1 eV and 0.6 eV) instead of the broad peak feature attributed to oxygen from angle-resolved photoemission spectra.~\cite{Shen1990,shen1991} \editt{This result is also at odds with
the results of DFT+DMFT calculations.~\cite{mandal2019,kunes2007}} The width of the valence band Ni $3d$ spectrum is also narrower overall than might be expected, as well as appearing sharper for the aforementioned reasons. However, the same experiments also indicate that the NiO valence-band spectrum has two distinct peaks, one around -2 eV and the other around -5 eV, a behavior reflected in our own plots, where the peak-to-peak splitting across the Fermi energy is approximately 4.6 eV, a value comparable to that derived from experiment (4.3 eV)\cite{Sawatzky1984,Tjernberg1996} and DMFT.\cite{mandal2019,kunes2007} We refer the reader to the SM for an analogous figure to Figure~\ref{Superexchange_Schematic}, where instead the 
$U^d$ is varied (SM Figure 2), as well as for an indicative plot of the total DOS (SM Figure 3) with no PAW sphere truncation, for the best performing functional that used U$^d_\textrm{eff}=5.11$ eV.} \editt{SM Figure 3, in particular, displays more clearly the low-energy peak at -5.1 eV of primarily Ni $e_g$ character, which has the distinct sharpness as the spectral feature attributed to Ni in experiment.}

\section{\label{SummaryConclusions} Summary and Conclusions}

In this work, we analyzed the NN and NNN Heisenberg exchange coupling coefficients of the prototypical Mott-Hubbard insulator NiO, calculating its electronic structure in its three main enforceable magnetic configurations (FM, AF$_\textrm{I}$, and AF$_\textrm{II}$) in a Hubbard corrected DFT context, for comparison with experiment. First, we examined the sensitivity of the first-principles-derived Hubbard \editt{parameters U and J$_\textrm{z}$}  to the PAW projectors. Here, we found that \texttt{dmatpuopt=3}, the population analysis corresponding to the once-normalized projection of the PAW basis functions on bound AE atomic eigenfunctions and truncated at the PAW augmentation radius, renders a numerically well-behaved linear response procedure that yields robust Hubbard parameters. These go on to provide experimentally relevant $J_1$ and $J_2$ exchange pairs once applied to NiO via a range of host Hubbard functionals.

Our results highlight just how valuable to the simulation of NiO and its total energetic properties are its in situ-derived \editt{U and  J$_\textrm{z}$}. We report rigorously defined PBE Hubbard \editt{parameters U and  J$_\textrm{z}$} for the Ni \textit{3d} and O \textit{2p} subspaces of NiO. The technical platform on which we build these investigations, the open-source plane-wave DFT suite \textsc{Abinit},\cite{gonze_abinit_2009,Amadon2008} allowed us to at once assess the numerical performance of the code's linear-response implementation for calculating the Hubbard U and to extend its utility to also calculate the \editt{intra-atomic} exchange coupling \editt{corrective parameter J$_\textrm{z}$} from first principles.

We continued to probe the minutiae of population analysis procedures to reevaluate spin-moment parametrization conventions, both inter- and intra-atomic, in the Heisenberg model. In doing so, we systematically charted a particular region of DFT-accessible spin-moment parameter space to make navigation of these techniques more manageable. Our optimization of this parameter space corresponds to the particular combination of (i) a rudimentary-quantum, atomic-sphere-restricted intra-atomic spin moment ($| \textbf{S}^\textrm{AS} |=\sqrt{2}\  \textrm{S}^\textrm{AS}_z $), and (ii) a  spin-parametrization scheme that assumes $\vert\textrm{S}_\textrm{FM}\vert=\vert\textrm{S}_{\textrm{AF}_{\textrm{I}}}\vert=\vert\textrm{S}_{\textrm{AF}_{\textrm{II}}}\vert$. We recommend this particular spin treatment for magnetic properties and total energy differences pertaining to NiO, given that the resulting $J_1$,$J_2$ exchange pairs give  \% RMS errors (with respect to experiment) as low as 13\%, a considerable improvement on state-of-the-art DFT simulations conducted thus far. Regardless of the Hubbard functional used, this spin treatment produced exchange coupling constants that are satisfactorily in line with experiment and highly competitive with other, more computationally taxing hybrid functionals, which were methodically categorized in a comprehensive literature review.

Throughout these investigations are implicit suggestions towards best practices in calculating Hubbard parameters and use of Hubbard functionals for non-ground state magnetic orderings of TMOs. Notably, we highlight the fact that the Hubbard functionals tested here, despite the bespoke nature of these parameters to their respective magnetic environments, did not result in comparable total energies, so we generally disadvise their use for such comparisons in NiO. Alternatively, these results highlight the need for more advanced Hubbard functionals designed to accommodate differences in magnetic environment. Notwithstanding, it is clear from Table \ref{FunctionalsVsExchange} that use of the \editt{J$_\textrm{z}$} parameter on both the Ni and the O is critical for reducing \% RMSE of the Heisenberg exchange coupling parameters in NiO with standard functionals.

Lastly, through extensive testing of a suite of 6+ different corrective Hubbard functionals, we draw  attention to the effect of the O \textit{2p} Hubbard parameter pairs on the NiO exchange coupling coefficients and provide a justification for the phenomenon thereabouts based on the crystal-field resolved DOS and occupation analyses.

\section{\label{Acknowledgements} Acknowledgements}

This research was supported by the Trinity College Dublin Provost PhD Project Awards. 
D.O'R. acknowledges the support of Science Foundation Ireland (SFI) [19/EPSRC/3605 \& 12/RC/2278\_2] and the Engineering and Physical Sciences Research Council [EP/S030263/1].
This work was further supported by Research IT at Trinity College Dublin. Calculations were principally performed on both the Boyle and Lonsdale clusters maintained by Research IT at Trinity College Dublin, the former of which was funded through grants from the European Research Council and Science Foundation Ireland, the latter through grants from Science Foundation Ireland. Further computational resources, facilities, and support were provided by the Irish Centre for High-End Computing (ICHEC). The authors would like to thank Daniel Lambert, Andrew Burgess, Stefano Sanvito, and Alessandro Lunghi for helpful discussions. 

\section{\label{Appendix} Appendix}
\appendix

\section{\label{Chi_0_Equivalency} Equivalency of Unscreened Response Functions for \texorpdfstring{$\alpha$}{alpha} and \texorpdfstring{$\beta$}{beta} Perturbations}

It is important to note, for verification purposes, that the unscreened response matrices $\chi_0$ and $\chi_{\textrm{M}_0}$ are equivalent. The proof of this is as follows.

Suppose we apply, to separate spin channels, an identical potential perturbation $dV^\uparrow_{ext}=dV^\downarrow_{ext}=dV_{ext}=d\alpha$ to a subspace on an atom. The unscreened change in spin-up occupancy $n_0^\uparrow$ as a result of this perturbation is
\begin{align}\label{proof0}
\frac{dn_0^\uparrow}{dV_{ext}}=\frac{dn_0^\uparrow}{dV_{ext}^\uparrow}\frac{dV_{ext}^\uparrow}{dV_{ext}}+\frac{dn_0^\uparrow}{dV_{ext}^\downarrow}\frac{dV_{ext}^\downarrow}{dV_{ext}}.
\end{align}

The second term of Eq. (\ref{proof0}) disappears because the occupancy of a spin channel does not change as a result of a potential perturbation on the opposing spin channel (i.e., $dn_0^\uparrow/dV_{ext}^\downarrow=0$). Then $dV_{ext}^\uparrow/dV_{ext}=1$, leaving  
\begin{align}\label{proof1}
\frac{dn_0^\uparrow}{dV_{ext}}=\frac{dn_0^\uparrow}{dV_{ext}^\uparrow}=\frac{dn_0^\uparrow}{d\alpha}.
\end{align}

Similarly, via the same procedure, we find that
\begin{align}\label{proof2}
\frac{dn_0^\downarrow}{dV_{ext}}=\frac{dn_0^\downarrow}{d\alpha}
\end{align}
and thus the unscreened change in total occupancy of the subspace, where $N_0=n_0^\uparrow+n_0^\downarrow$, is
\begin{align}\label{proof3}
\chi_0&=\frac{dN_0}{dV_{ext}}=\frac{d(n_0^\uparrow+n_0^\downarrow)}{d\alpha}.
\end{align}

Now consider that on a completely identical subspace on an identical atom, we apply a potential perturbation of $dV^\uparrow_{ext}=\beta$ to the spin up channel and $dV^\downarrow_{ext}=-\beta$ to the spin down channel. Here, we observe the unscreened change in magnetization $\mu_0=n_0^\uparrow-n_0^\downarrow$ as a result of this perturbation, which is
\begin{align}\label{proof4}
\chi_{\textrm{M}_0}&=\frac{d\mu_0}{dV_{ext}}=\frac{d(n_0^\uparrow-n_0^\downarrow)}{d\beta} \nonumber \\
&=\frac{dn_0^\uparrow}{d\beta}-\frac{dn_0^\downarrow}{d\beta}.
\end{align}

The change in spin-up occupancy with respect to $\beta$ is found to be identical to Eq. (\ref{proof0}). But because a $-\beta$ perturbation is applied to the spin down channel potential,
\begin{align}\label{proof5}
\frac{dn_0^\downarrow}{dV_{ext}}=-\frac{dn_0^\downarrow}{d\beta}.
\end{align}

Thus,
\begin{align}\label{proof6}
\chi_{\textrm{M}_0}&=\frac{d\mu_0}{dV_{ext}}=\frac{dn_0^\uparrow}{d\beta}-\left(-\frac{dn_0^\downarrow}{d\beta}\right) \nonumber \\
&\therefore \chi_{\textrm{M}_0}=\frac{d(n_0^\uparrow+n_0^\downarrow)}{d\beta}
\end{align}

In the event that $\alpha=\beta$, therefore, $\chi_0=\chi_{\textrm{M}_0}$, a conclusion following Eqs. (\ref{proof3}) and (\ref{proof6}).

\bibliography{./Manuscript}

\end{document}

\title{Supplemental Material for:\\
Optimization strategies developed on NiO for Heisenberg exchange coupling
calculations using projector augmented wave based first-principles DFT+U+J}

\author{L. MacEnulty}
\author{D. D. O'Regan}
\affiliation{
 School of Physics, Trinity College Dublin,\\
 The University of Dublin, Dublin 2, Ireland\\
}

\date{\today}

\maketitle

\section{\label{SM:Spin Moment Method Tables}Data concerning various spin moment techniques}

We report here the data pertaining to the $\textrm{S}_z$ and $|\textbf{S}|$
calculations across all three population analysis methods for the spin moment: the atomic spheres (AS) method of Eq. (24) of the manuscript, the PAW occupations (PAW) of Eq. (25), and the full spin density (SD) as embodied by Eq. (26). Table V of the main work is a condensed reflection of this table. To acquire the numbers in Table V, we take the cumulative \% RMS error of the exchange parameter pairs with respect to the Hutchings \textit{et al}\cite{Hutchings_1972} experimentally obtained parameters, then average across all six Hubbard functionals (i.e., excluding the raw PBE data).

\newcolumntype{?}{!{\vrule width 2pt}}
\begin{table*}[!h]
\begin{tabular}{cr?cc|cc|cc?cc|cc|cc|}
\multicolumn{1}{l}{} & \multicolumn{1}{l}{} & \multicolumn{6}{c}{$\textrm{S}_z$} & \multicolumn{6}{c}{$|\textbf{S}|$} \\ \cmidrule(lr){3-8}\cmidrule(lr){9-14}

\multicolumn{1}{c}{{\color[HTML]{FFFFFF} }} & \multicolumn{1}{c}{{\color[HTML]{FFFFFF} }} & \multicolumn{2}{c}{\color{tan} Method A} & \multicolumn{2}{c}{\color{navy} Method B} & \multicolumn{2}{c?}{\color{copper} Method C} & \multicolumn{2}{c}{\color{tan} Method A} & \multicolumn{2}{c}{\color{navy} Method B} & \multicolumn{2}{c}{\color{copper} Method C} \\ \cmidrule(lr){3-4}\cmidrule(lr){5-6}\cmidrule(lr){7-8}\cmidrule(lr){9-10}\cmidrule(lr){11-12}\cmidrule(lr){13-14}

\rowcolor[HTML]{FFFFFF}[\overhang]
\multicolumn{1}{c}{{\color[HTML]{FFFFFF} }} & \multicolumn{1}{c?}{{\color[HTML]{FFFFFF} }} & {\tcw\cellcolor{tan} $J_1$} & {\tcw\cellcolor{tan} $J_2$} & {\tcw\cellcolor{navy} $J_1$} & {\tcw\cellcolor{navy} $J_2$} & {\tcw\cellcolor{copper} $J_1$} & {\tcw\cellcolor{copper} $J_2$} & {\tcw\cellcolor{tan} $J_1$} & {\tcw\cellcolor{tan} $J_2$} & {\tcw\cellcolor{navy} $J_1$} & {\tcw\cellcolor{navy} $J_2$} & {\tcw\cellcolor{copper} $J_1$} & {\tcw\cellcolor{copper} $J_2$} \\

\rowcolor[HTML]{FFFFFF}[\overhang]
\multicolumn{1}{c}{} & \multicolumn{1}{r?}{\textbf{Functional\spc}} & {\tcw\cellcolor{tan} (meV)} & {\tcw\cellcolor{tan}(meV)} & {\tcw\cellcolor{navy} (meV)} & {\tcw\cellcolor{navy}(meV)} & {\tcw\cellcolor{copper} (meV)} & {\tcw\cellcolor{copper}(meV)} & {\tcw\cellcolor{tan} (meV)} & {\tcw\cellcolor{tan}(meV)} & {\tcw\cellcolor{navy} (meV)} & {\tcw\cellcolor{navy}(meV)} & {\tcw\cellcolor{copper} (meV)} & {\tcw\cellcolor{copper}(meV)} \\ \hline

\multirow{7}{*}{\textbf{AS}} & \multicolumn{1}{r?}{DFT (PBE)} & 8.23 & -29.32 & 19.46 & -69.35 & 2.74 & -60.96 & 8.23 & -29.32 & 9.73 & -34.68 & 1.37 & -30.48 \\
 & DFT+U$^{d}$ & 0.95 & -10.20 & 1.34 & -14.43 & 1.20 & -13.91 & 0.95 & -10.20 & 0.67 & -7.22 & 0.60 & -6.96 \\
 & DFT+U$^{d,p}$ & 1.53 & -7.11 & 2.15 & -10.00 & 2.00 & -9.76 & 1.53 & -7.11 & 1.08 & -5.00 & 1.00 & -4.88 \\
 & DFT+U$^{d}_\textrm{eff}$ & 1.03 & -11.10 & 1.48 & -16.00 & 1.32 & -15.36 & 1.03 & -11.10 & 0.74 & -8.00 & 0.66 & -7.68 \\
 & DFT+U$^{d,p}_\textrm{eff}$ & 1.52 & -8.41 & 2.19 & -12.07 & 2.00 & -11.72 & 1.52 & -8.41 & 1.10 & -6.04 & 1.00 & -5.86 \\
 & DFT+U$^{d}$ + J$^{d}_{\textrm{z}}$ & 0.96 & -10.24 & 1.37 & -14.61 & 1.23 & -14.08 & 0.96 & -10.24 & 0.69 & -7.31 & 0.62 & -7.04 \\
 & DFT+U$^{d,p}$ + J$^{d,p}_{\textrm{z}}$ & 1.00 & -7.35 & 1.43 & -10.44 & 1.32 & -10.19 & 1.00 & -7.35 & 0.72 & -5.22 & 0.66 & -5.10 \\ \hline

\multirow{7}{*}{\textbf{PAW}} & \multicolumn{1}{r?}{DFT (PBE)} & 8.23 & -29.32 & 19.00 & -67.70 & 3.17 & -60.58 & 8.23 & -29.32 & 9.50 & -33.85 & 1.59 & -30.29 \\
 & DFT+U$^{d}$ & 0.95 & -10.20 & 1.28 & -13.74 & 1.19 & -13.40 & 0.95 & -10.20 & 0.64 & -6.87 & 0.60 & -6.70 \\
 & DFT+U$^{d,p}$ & 1.53 & -7.11 & 2.05 & -9.52 & 1.95 & -9.35 & 1.53 & -7.11 & 1.03 & -4.76 & 0.98 & -4.68 \\
 & DFT+U$^{d}_\textrm{eff}$ & 1.03 & -11.10 & 1.41 & -15.26 & 1.31 & -14.84 & 1.03 & -11.10 & 0.71 & -7.63 & 0.66 & -7.42 \\
 & DFT+U$^{d,p}_\textrm{eff}$ & 1.52 & -8.41 & 2.09 & -11.52 & 1.96 & -11.27 & 1.52 & -8.41 & 1.05 & -5.76 & 0.98 & -5.64 \\
 & DFT+U$^{d}$ + J$^{d}_{\textrm{z}}$ & 0.96 & -10.24 & 1.31 & -13.99 & 1.23 & -13.64 & 0.96 & -10.24 & 0.66 & -7.00 & 0.62 & -6.82 \\
 & DFT+U$^{d,p}$ + J$^{d,p}_{\textrm{z}}$ & 1.00 & -7.35 & 1.36 & -10.00 & 1.29 & -9.81 & 1.00 & -7.35 & 0.68 & -5.00 & 0.65 & -4.91 \\ \hline

\multirow{7}{*}{\textbf{SD}} & \multicolumn{1}{r?}{DFT (PBE)} & 8.23 & -29.32 & 15.89 & -56.63 & 3.13 & -46.94 & 8.23 & -29.32 & 7.95 & -28.32 & 1.57 & -23.47 \\
 & DFT+U$^{d}$ & 0.95 & -10.20 & 1.15 & -12.33 & 1.46 & -11.66 & 0.95 & -10.20 & 0.58 & -6.17 & 0.73 & -5.83 \\
 & DFT+U$^{d,p}$ & 1.53 & -7.11 & 1.84 & -8.55 & 1.93 & -8.02 & 1.53 & -7.11 & 0.92 & -4.28 & 0.97 & -4.01 \\
 & DFT+U$^{d}_\textrm{eff}$ & 1.03 & -11.10 & 1.26 & -13.62 & 1.61 & -12.80 & 1.03 & -11.10 & 0.63 & -6.81 & 0.81 & -6.40 \\
 & DFT+U$^{d,p}_\textrm{eff}$ & 1.52 & -8.41 & 1.83 & -10.13 & 2.00 & -9.52 & 1.52 & -8.41 & 0.92 & -5.07 & 1.00 & -4.76 \\
 & DFT+U$^{d}$ + J$^{d}_{\textrm{z}}$ & 0.96 & -10.24 & 1.17 & -12.51 & 1.48 & -11.80 & 0.96 & -10.24 & 0.59 & -6.26 & 0.74 & -5.90 \\
 & DFT+U$^{d,p}$ + J$^{d,p}_{\textrm{z}}$ & 1.00 & -7.35 & 1.22 & -8.95 & 1.44 & -8.38 & 1.00 & -7.35 & 0.61 & -4.48 & 0.72 & -4.19 \\ \hline
 
\end{tabular}
\caption{Comparison of exchange coupling parameters from various Hubbard functionals employing the \emph{AF$_\textrm{II}$} HPs in column 4 of Table IV (a $^d$ superscript refers to parameter on \emph{Ni} 3d orbitals; $^p$ superscript refers to parameter on \emph{O} 2p orbitals; \emph{U$_\textrm{eff}=\textrm{U}-\textrm{J}_\textrm{z}$}). Moments from \emph{SD} harvested from Abinit output using C2x.\cite{rutter_c2x_2018}}
\label{AS_SD_PAW_Sz_S_Table}
\end{table*}

\newpage

\section{\label{SM:UdDOS}Density of States plot for \texorpdfstring{DFT+U$^d$}{DFT+Ud}}

%%%%%%%%%%%%%%%%%%%%%%%% Ud DOS %%%%%%%%%%%%%%%%%%%%%%%%%%
\begin{figure*}[!h]
    \centering
    \includegraphics[width=1.0\textwidth]{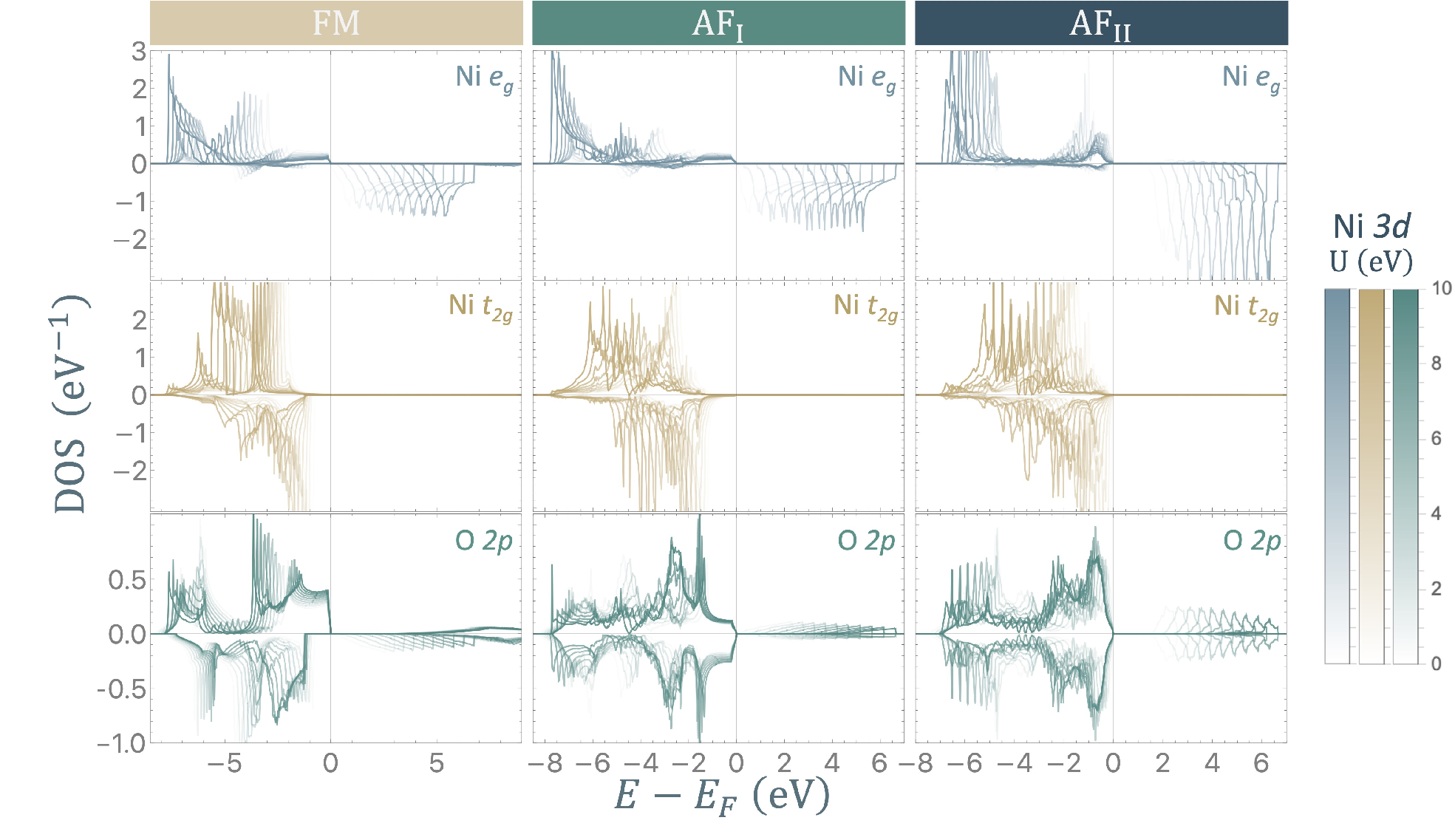}
    \caption{\edit{Overlaid atomic (AE) orbital projected density of states (PDOS) on two particular atoms of the four-atom unit cell, resolved by crystal field splitting characterization on \emph{Ni} (e$_g$ blue, top row; t$_{2g}$ tan, middle row) and \emph{O} (\textit{2p} orbital green, bottom row) for each magnetic ordering (\emph{FM} left column; \emph{AF$_\textrm{I}$} middle column; \emph{AF$_\textrm{II}$} right column) of \emph{NiO}. Positive values indicate majority spin PDOS, negative values indicate minority spin. Opacity of plot is a function of \emph{U$^d$ (eV)} in the Dudarev \emph{DFT+U$^{d}$} functional, where U$^{d}$ is applied in the same magnitude to all MOs of NiO. Only the component within the PAW augmentation sphere cutoff $r_c$ for each species is included, with no interstitial contributions, hence the relatively sharp (atomic-like) features of the PDOS. Moreover, only the conduction band edge position should be interpreted here, as the Kohn-Sham basis is not thought to be complete for the conduction band for all MOs in the energy range shown.}}
    
    \label{UpVariation}
\end{figure*}
%%%%%%%%%%%%%%%%%%%%%%%%%%%%%%%%%%%%%%%%%%%%%%%%%%%%%%%%%%

\newpage

\section{\label{SM:ConventionalDOSsection}Unit-Cell totalled Density of States plot for \texorpdfstring{DFT+U$^d$}{DFT+Ud=5.11 eV}}

%%%%%%%%%%%%%%%%%% Conventional DOS %%%%%%%%%%%%%%%%%%%%%%%
\begin{figure*}[!h]
    \centering
    \includegraphics[width=1.0\textwidth]{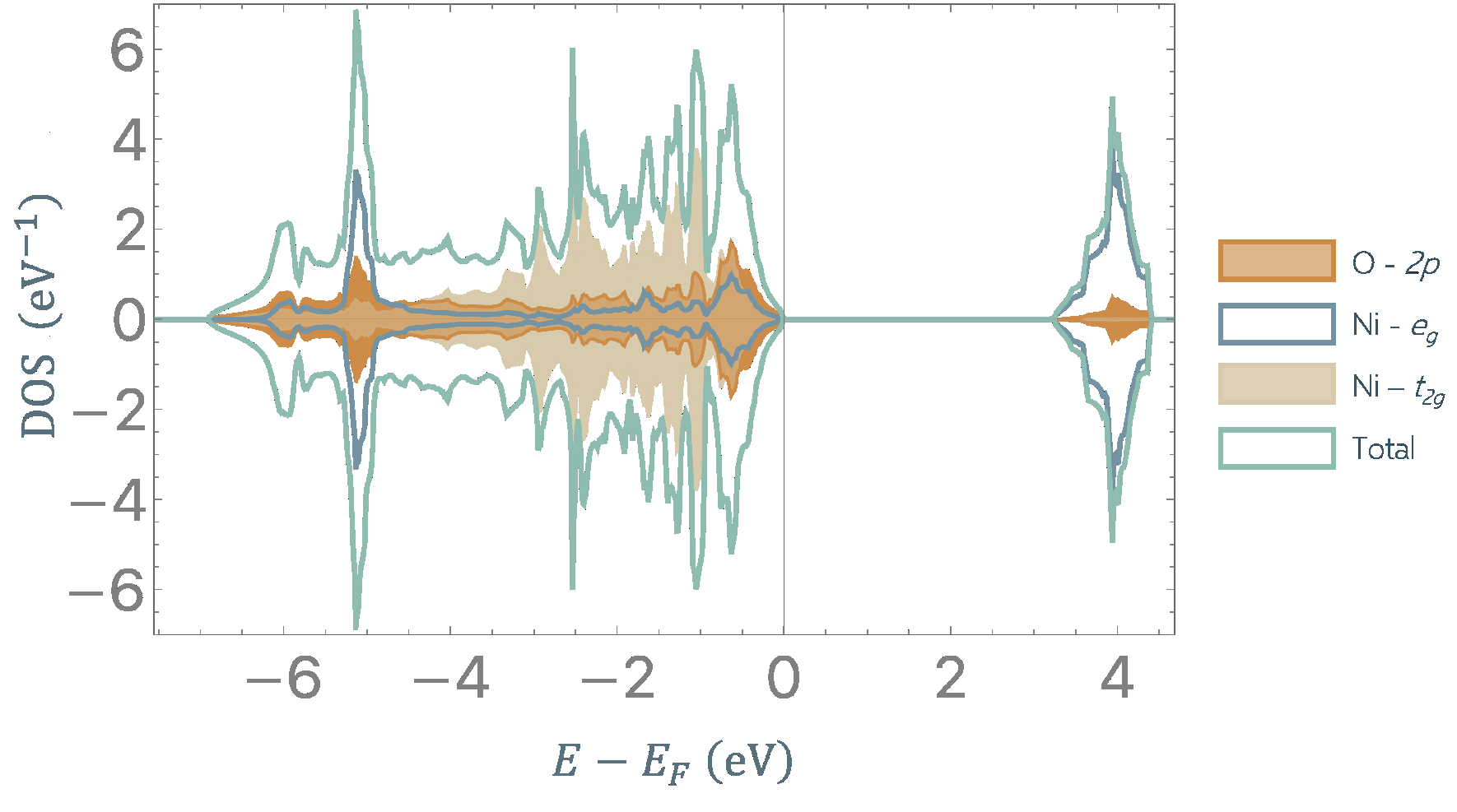}
    \caption{\edit{PDOS resolved by crystal field splitting characterization, summed over all atoms in the unit cell, \emph{Ni} (e$_g$ blue; t$_{2g}$ tan) and \emph{O} (\textit{2p} orange)  of \emph{NiO}. Calculated with the Dudarev functional U$^d_\textrm{eff}=5.11$ eV. Only the component within the PAW augmentation sphere cutoff $r_c$ for each species is included, with no interstitial contributions, hence the relatively sharp (atomic-like) features of the PDOS. The Total DOS (light green) is also shown, which is not restricted to the PAW augmentation sphere region and does include all contributions. In all cases, only the conduction band edge position should be interpreted here, as the Kohn-Sham basis is not thought to be complete for the conduction band for all MOs in the energy range shown.}}
    \label{SM:ConventionalDOS}
\end{figure*}
%%%%%%%%%%%%%%%%%%%%%%%%%%%%%%%%%%%%%%%%%%%%%%%%%%%%%%%%%%

\FloatBarrier

\providecommand{\noopsort}[1]{}\providecommand{\singleletter}[1]{#1}